\documentclass[12pt,1p,authoryear]{elsarticle}
\usepackage{snapshot}
\usepackage{a4wide}
\usepackage[utf8]{inputenc}
\usepackage[english]{babel}
\usepackage{booktabs}
\usepackage{rotating}
\usepackage[names]{xcolor}
\usepackage{colortbl}
\usepackage{multirow}
\definecolor{tableShade}{gray}{0.97}
\definecolor{col1}{gray}{0.97}
\definecolor{col2}{gray}{1}

\usepackage{tikz}
\usepackage{amsthm,amssymb,amsmath,latexsym}
\usepackage{verbatim}
\usepackage{bm}

\newcommand{\wAIC}{\ensuremath{\operatorname{AIC}_{\text{\tiny IPCW}}}}
\newcommand{\cause}{\epsilon}
\newcommand{\ocause}{\widetilde{\cause}}
\newcommand{\otime}{\widetilde{T}}
\newcommand{\latent}{\eta}

\newcommand{\var}{\ensuremath{\mathbb{V}\text{ar}}} 
\newcommand{\cov}{\ensuremath{\mathbb{C}\text{ov}}} 
 
\newcommand{\E}{\ensuremath{\mathbb{E}}} 
\newcommand{\pr}{\ensuremath{\mathbb{P}}}

\newcommand{\Gc}{G_c}
\newcommand{\hatGc}{\widehat{G}_c}
\newcommand{\transpose}{T}
 
\newcommand{\diag}{\operatorname{diag}} 
 
\newcommand{\independenT}[2]{\mathrel{\setbox0\hbox{$#1#2$}\copy0\kern-\wd0\mkern4mu\box0}} 
 
\newcommand{\influ}{\ensuremath{I\!F}}
\newtheorem{thm}{Theorem}

\begin{document}

\begin{frontmatter}

\title{The Liability Threshold Model for Censored Twin Data}

\author[biostatku]{Klaus K.~Holst\corref{cor1}}
\ead{k.k.holst@biostat.ku.dk}
%%\address{ Øster Farimagsgade 5, 1014K Copenhagen, Denmark}
\cortext[cor1]{Corresponding author. Tel.: +45 35327901}
\author[biostatku]{Thomas H.~Scheike}
\author[biostatsdu]{Jacob B.~Hjelmborg}
\address[biostatku]{Dept.~of Biostatistics, University of Copenhagen, Denmark}
\address[biostatsdu]{Dept.~of Epidemiology, Biostatistics and Biodemography, University of Southern Denmark, Denmark}

\begin{abstract}

  Family studies provide an important tool for understanding etiology
  of diseases, with the key aim of discovering evidence of family
  aggregation and to determine if such aggregation can be attributed
  to genetic components.  Heritability and concordance estimates are
  routinely calculated in twin studies of diseases, as a way of
  quantifying such genetic contribution. The endpoint in these studies
  are typically defined as occurrence of a disease versus death
  without the disease. However, a large fraction of the subjects may
  still be alive at the time of follow-up without having experienced
  the disease thus still being at risk.  Ignoring this right-censoring
  can lead to severely biased estimates.  The classical liability
  threshold model can be extended with inverse probability of
  censoring weighting of complete observations.  This leads to a
  flexible way of modelling twin concordance and obtaining consistent
  estimates of heritability. The method is demonstrated in simulations
  and applied to data from the population based Danish twin cohort to
  describe the dependence in prostate cancer occurrence in twins.

  % The method corrects for a major source of bias by taking advantage
  % of the time to event information that is most often provided in
  % cohort studies along with the binary trait.
  
  % We extend classical methodology for studying genetic influence of a
  % binary trait observed in twin pairs with missing data.
  % estimation-ipcw-simulation-features-implementation-application
  % given that we have observed time to event or to
  % follow-up.   
  % providing cumulative concordance and heritability by age.  
\end{abstract}


\end{frontmatter}

%}}} frontmatter

%{{{ Introduction

\section{Introduction}\label{sec:intro}
Family studies provide an important tool for understanding etiology of
diseases, with the key aim of discovering evidence of family
aggregation and to determine if such aggregation can be attributed to
genetic components.  Heritability and concordance estimates are
routinely calculated in twin studies of diseases, as a way of
quantifying such genetic contribution.  As a key paper for studying
heritability of cancer, \cite{lichtenstein2000environmental} reported
heritability estimates for prostate cancer of 0.42 (95\% confidence
limits 0.29--0.50) and casewise concordance of 0.21 in monozygotic
(MZ) twins and 0.06 in dizygotic (DZ) twins based on combined cohorts
of 44,788 twin pairs from the Nordic twin registries.  This suggests a
considerable genetic contribution to the development of prostate
cancer.  A polygenic liability threshold model, i.e., a Probit
variance component model, was used to quantify the heritability on the
liability scale from the classification of subjects as cancer cases or
non-cancer cases (died without cancer). However, a large fraction of
the twin-pairs were still alive at the end of follow-up but treated as
non-cancer case. This corresponds to treating this part of the
population as immune to cancer, suggesting that the estimates of the
targeted population parameters in this study could be severely
biased. The censoring mechanism has largely been ignored in the
epidemiological literature of family studies, which unfortunately
makes reported estimates of both heritability, and other population
parameters of interest such as concordance probabilities, very
difficult to interpret.

The key to solving this problem is to consider the event times in the
analysis. Standard techniques for correlated survival data are not
appropriate here, due to the competing risk of death.  Dependence on
the hazard scale while taking possible dependence between causes into
account has been considered by \cite{ripatti2003} and
\cite{gorfinehsu2011}. \cite{scheike13:lida} considered dependence on
the probability scale via random effects models and
\cite{scheike13:concordance} examined non-parametric estimates of the
concordance function, i.e., the probability of both twins experiencing
cancer before a given time point.  These methods yield constructive
ways of analysing twin data of disease status, however, care in
correctly specifying the dependence structure over time via the random
effects structure has to be taken. Furthermore, none of the approaches
provide heritability estimates that are comparable with the classical
definition of heritability on the liability scale given by
\cite{falconer67}.  In the following we will define a simple estimator
which gives consistent concordance estimates and estimates of
heritability on the liability scale under independent right-censoring.

% motivate the approach, model with extension.
% The liability threshold model analyses have been criticized for
% completely ignoring the time-aspect and the fact that
%%%Probit random effects
%%%a large number of subjects are
%%%censored.  In addition
% the analyses did not take the censoring into account which makes it
% impossible to interpret the results, as indicated above.

\vspace*{\bigskipamount}

The paper is structured as follows. In Section~\ref{sec:genetics} we
review basic concepts in quantitative genetics and define heritability
with the aim of estimating the degree of association due to genes and
environmental factors through random effects modelling. In particular,
we note that dependence on the probability scale is something quite
different from dependence on the normal scale. We introduce the
competing risks framework and present the inverse probability of
censoring weighted estimating equations in Section~\ref{sec:ipcw}. The
method is demonstrated in simulations in Section~\ref{sec:sim}.
% more specific details on
% how to do regression modelling for censored data that quantifies
A worked example based on the Danish twin registry is presented in
Section~\ref{sec:application} followed by a general discussion.

%}}} Introduction

%{{{ Quantitative genetics

\section{Polygenic models}\label{sec:genetics}

The basic idea of family-studies of a quantitative trait is to exploit
that stronger phenotypic resemblance will be seen between closely
related family members when the trait is genetically determined.  In
particular, for twin studies we may exploit that monozygotic (MZ)
twins in principle are genetic copies whereas dizygotic (DZ) twins
genetically on average resembles ordinary siblings.  This allows us
under appropriate genetic assumptions to decompose the trait into
genetic and environmental components, $Y=Y_{\text{gene}} +
Y_{\text{envir}}$, which may be modelled using random effects.
Assuming independence between genetic and environmental effects the
\textit{broad-sense heritability} may then be quantified as the
fraction of the total variance due to genetic factors.

The theoretic foundation in modern quantitative genetics was laid out
in the pioneering work of \cite{Fisher:1918} who formally described
the above genetic decomposition in terms of additive and dominant
genetic effects. Familial resemblance may defined from the
\textit{kinship-coefficient} $\Phi_{jk}$ which is the probability that
two randomly selected alleles from the same locus of relatives $k$ and
$j$ are \textit{identical by descent}, i.e., the alleles are physical
copies of the same gene carried by a common ancestor.  Under
assumptions of random mating (no inbreeding), linkage equilibrium, no
gene-environment interaction and epistasis, and parents do not
transmit their environmental effects to their children, this leads to
a covariance between the observed phenotypes $Y_k$ and $Y_j$ for the
relatives given by
\begin{align*}
  \cov(Y_k,Y_j) = 2\Phi_{kj}\sigma_A^2 + \Delta_{7kj}\sigma_D^2 + \sigma_C^2,
\end{align*}
where the identity coefficient $\Delta_{7kj}$ describes the
probability that at a given loci both alleles for the two relatives
are identical by descent
\citep{lange02:_mathem_statis_method_genet_analy}. The variance
components $\sigma_A^2$ describes the additive genetic effects,
$\sigma_D^2$ the dominant genetic effects and $\sigma_C^2$ describes
variance of shared environmental effects for the two relatives.

This can be captured in a random effects model where the polygenic phenotype
$Y_{ij}$ may be modelled as
\begin{align}\label{eq:polygenic1}
  Y_{ij} = \beta^\transpose X_{ij} + \latent^A_{ij} + \latent^C_{i} + \latent^D_{ij} + \varepsilon_{ij},
\end{align}
for family $i=1,\ldots,n$ and family member $j=1,\ldots,K$ with
covariates $X_{ij}$. Here we assume that there is the same shared
environmental effect for all family members. All the random effects
are assumed to be independent and normally distributed which in
general may be reasonable for polygenic traits \citep{lange97}
\begin{align*}
  (\latent^A_{ij}, 
    \latent^C_{i},
    \latent^D_{ij},
    \varepsilon_{ij})^\transpose
    \sim \mathcal{N}\left(0,\diag(\sigma_A^2,\sigma_C^2,\sigma_D^2,\sigma_E^2)\right).
\end{align*}
The residual terms $\varepsilon_{ij}$ are assumed to be iid normal and
the variance component $\sigma_E^2$ may be interpreted as the variance
of the unique environmental effects. The (broad-sense) heritability may
then be defined as
\begin{align*}
  H^2 =
    \frac{\sigma_A^2+\sigma_D^2}{\sigma_A^2+\sigma_C^2+\sigma_D^2+\sigma_E^2}.
\end{align*}
For MZ twins we have $\Phi_{kj}^{\text{MZ}}=\tfrac{1}{2}$ and
$\Delta_{7kj}^{\text{MZ}}=1$ and for DZ twins
$\Phi_{kj}^{\text{DZ}}=\Delta_{7kj}^{\text{DZ}}=\tfrac{1}{4}$, hence
\begin{align*}
  \cov(Y_{i1}^{\text{MZ}}, Y_{i2}^{\text{MZ}}) &=
  \begin{pmatrix}
    \sigma_A^2+\sigma_C^2+\sigma_D^2 + \sigma_E^2 & \sigma_A^2+\sigma_C^2+\sigma_D^2 \\
    \sigma_A^2+\sigma_C^2+\sigma_D^2 & \sigma_A^2+\sigma_C^2+\sigma_D^2 + \sigma_E^2
  \end{pmatrix}, \\
  \cov(Y_{i1}^{\text{DZ}}, Y_{i2}^{\text{DZ}}) &=
  \begin{pmatrix}
    \sigma_A^2+\sigma_C^2+\sigma_D^2+\sigma_E^2 & \tfrac{1}{2}\sigma_A^2+\sigma_C^2+\tfrac{1}{4}\sigma_D^2 \\
    \tfrac{1}{2}\sigma_A^2+\sigma_C^2+\tfrac{1}{4}\sigma_D^2 & \sigma_A^2+\sigma_C^2+\sigma_D^2 + \sigma_E^2
  \end{pmatrix}.
\end{align*}
Note that one consequence of the model is that MZ and DZ twins follows
the same marginal distribution.  Unfortunately, the classic twin
design does not allow identification of all variance
components. Further inclusion of other family members or
twin-adoptives can remedy this problem, but may further complicate
assumptions regarding shared/non-shared environmental effects across
different family members. The pragmatic solution is typically to
report results from the most biologically relevant model, i.e., for
certain traits the shared environmental effect may be known to be
negligible, or to choose a sub-model based on some model selection
criterion \citep{Akaike73}. For the classical twin design omitting one
variance component in the above formulation (typically the dominant
genetic component, leading to the so-called ACE-model), the Maximum
Likelihood Estimates can be obtained using specialised software for
family studies \citep{metspackage} or any general Structural Equation
Model implementation.

\subsection{Liability threshold model}

For binary traits the classical polygenic model \eqref{eq:polygenic1}
may be extended by a model of the form
\begin{eqnarray}
g(\pr(Y_{ij}=1 \mid X_{ij},\latent_{ij}^{A},\latent_{i}^{C},\latent_{ij}^{D})) = \beta^\transpose X_{ij} +
\latent_{ij}^{A} + \latent_{i}^{C} + \latent_{ij}^{D}, \quad j=1,2,
\label{eq:probit1}
\end{eqnarray}
where $g$ is some link-function, $X_{ij}$ are possible covariates that
we wish to adjust for, and $\latent_{ij}^{A},\latent_{ij}^{C},\latent_{ij}^{D}$ are random
effects.
\begin{figure}[htbp]
  \centering
  \includegraphics[width=0.6\textwidth]{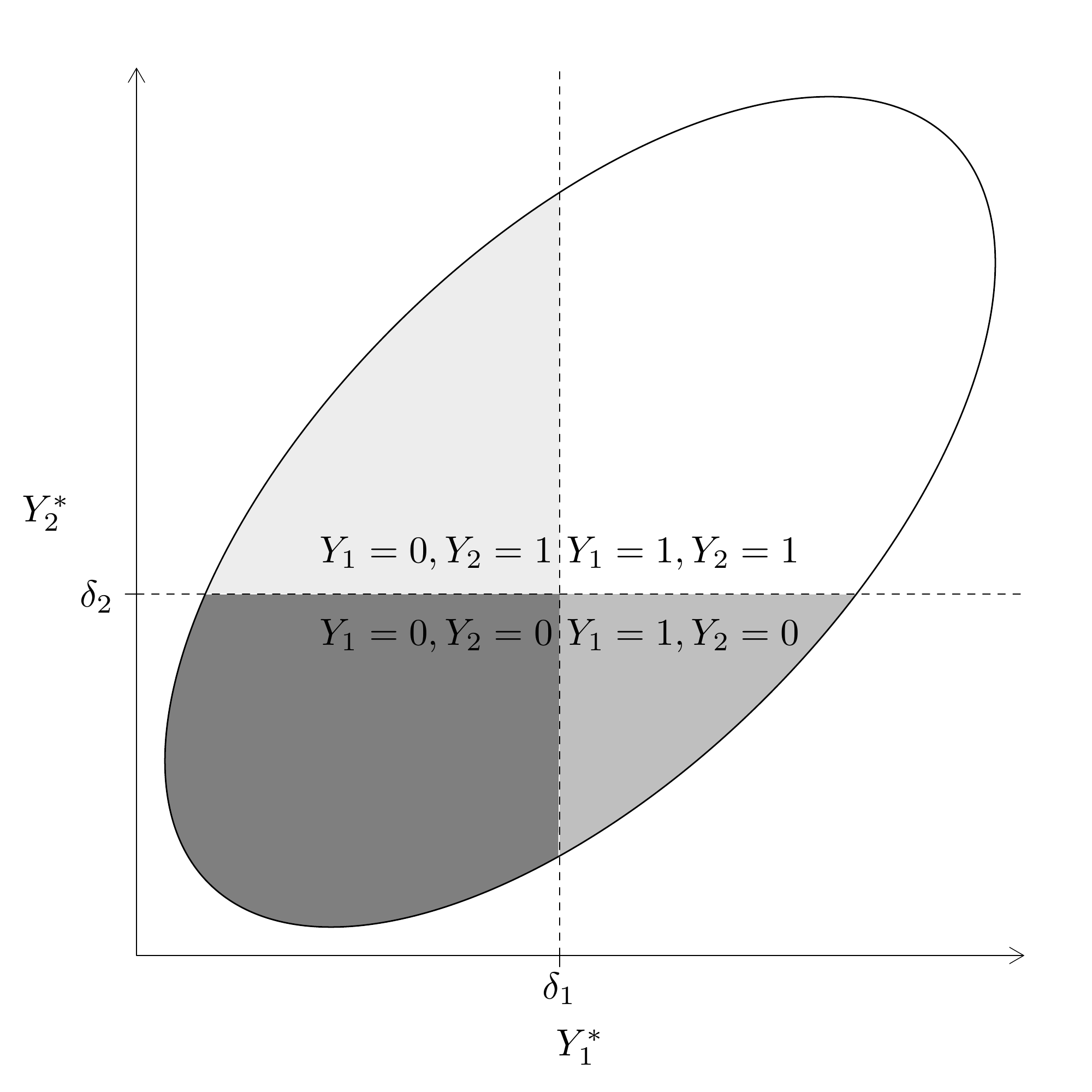}
  \caption{Liability threshold model where the observed binary pair
    $(Y_1,Y_2)$ is a realization defined from underlying unobserved
    continuous variables $(Y_1^*,Y_2^*)$ such that $Y_k=1$ exactly
    when the \textit{liability} $Y_k^*$ exceeds some threshold $\delta_k$.}
  \label{fig:threshold}
\end{figure}
Using the Probit link \citep{falconer67, falconer-mackay-1994,
  nealecardon, shamhuman} Equation \eqref{eq:probit1} gives the
\emph{Liability Threshold Model} and has been widely adopted, since
this leads to a model equivalent to \eqref{eq:polygenic1} for a latent
Gaussian variable (see Figure~\ref{fig:threshold})
\begin{align*}
  Y_{ij}^* = \beta^\transpose X_{ij} + \latent^A_{ij} + \latent^C_{i} +
  \latent^D_{ij} + \varepsilon_{ij}, \quad j=1,2,
\end{align*}
where we only observe the thresholded version
\begin{align*}
  Y_{ij} =
  \begin{cases}
    1,& Y_{ij}^*\geq\delta_j\\
    0,& Y_{ij}^*<\delta_j.
  \end{cases}
\end{align*}
For identification, the threshold is fixed at $\delta_j=0$
and the variance of the residual term $\epsilon_{ij}$ set to one.
On the Probit-scale this corresponds to 
\begin{align}\label{eq:liabilityprob}
  \pr(Y_{ij}=1 \mid X_{ij},\latent^A_{ij},\latent^C_{i},\latent^D_{ij}) =
  \Phi(\beta^\transpose X_{ij} +
  \latent^A_{ij} + \latent^C_{i} + \latent^D_{ij}),
\end{align}
noting that the E component is modelled indirectly through the
inverse link-function $\Phi$ which is the standard normal CDF,
i.e., $\sigma_E^2=1$.  In the following we will simplify notation and
use $\latent_{ij}$ to denote the total random effect for the $j$th
twin in the $i$th twin-pair.

Note that the corresponding heritability in this model
\begin{align*}
  H^2 = \frac{\sigma_A^2+\sigma_D^2}{\sigma_A^2+\sigma_C^2+\sigma_D^2+1},
\end{align*}
relates to the underlying liability scale, and that there are
additional variation present in the data on the risk scale.  Using
only the random effects to define a heritability estimate is thus not 
comparable to the one from the standard normal model, where all the
variation is included in the heritability estimate. 

The Probit random effects analyses have been criticized for completely
ignoring the time-aspect and the fact that the analyses did not take
censoring into account \citep{duncan04:genepi}.  In
\cite{lichtenstein2000environmental} the analysis was based on the
assumption that the probability of occurrence of cancer for twin $j$
in twin pair $i$ was on the same form as \eqref{eq:probit1} with
\begin{eqnarray}
  \pr(\text{twin $j$ gets cancer} \mid \latent_{ij}) = \Phi(\latent_{ij}), \quad j=1,2,
\label{eq:probit2}
\end{eqnarray}
and with the complementary outcome being that the twin died
without getting cancer or still was alive and without cancer at the
time of follow-up. The latter group are thus treated as immune to
cancer after they leave the study, which in general makes the results
of the analysis impossible to interpret. The right-censoring
mechanism therefore has to be taken into account, but
additional information on the timing of the events are needed.  In
practice, these event times are typically readily available in family
studies of disease.

\begin{figure}[htbp]
  \centering
\tikzstyle{plain2}=[rectangle,thick,minimum height=1.2cm,minimum width=2cm,draw=gray!80]
  \begin{tikzpicture}[>=latex,text height=1.5ex,text depth=0.25ex]
    \matrix[row sep=0.8cm,column sep=2cm]{ 
      & \node(D) [plain2] {Dead}; \\
      \node(A) [plain2] {Alive}; \\
      & \node(P) [plain2] {\parbox{1.4cm}{\hfuzz=8pt Prostate \\cancer}}; \\
    };
    \path[->] (A) edge[thick] node [auto] {$\alpha_{13}(t)$} (D) ;
    \path[->] (A) edge[thick] node [auto, swap] {$\alpha_{12}(t)$} (P) ;
    \path[->] (P) edge[thick] node [auto, swap] {$\alpha_{23}(t)$} (D) ;
  \end{tikzpicture}  
  \caption{Competing risks model for the two competing risks of death
    and prostate cancer with the transition probabilities being
    described by the cause-specific hazards
    $\alpha_{kl}(t)$.\label{fig:comprisk}}
\end{figure}
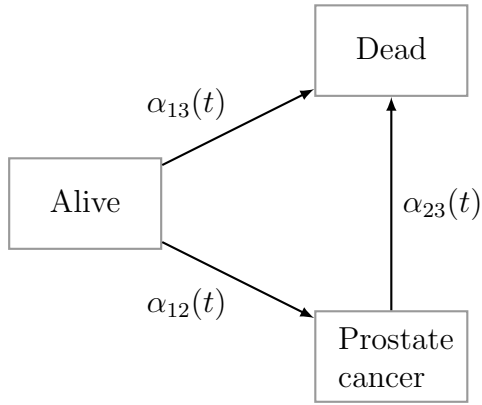

%}}} Quantitative genetics

%{{{ Comp.risk & IPCW

\section{Inverse Probability of Censoring Weighted Estimating Equation}\label{sec:ipcw}

%\subsection{Competing risks setup}
The definition of the liability threshold model perceives the states
``prostate cancer'' and ``death'' as static endpoints.  Our aim of
adjusting the estimating procedure for the right-censoring, however,
requires us to consider the data in a dynamic framework.  A more
natural setting for the data generating mechanism is to consider the
problem in the competing risk setting. In the following let $(T_{ik},
C_{ik},\cause_{ik},X_{ik})$ denote the event time, right censoring
time, the cause of failure $\cause_{ik}\in\{1,\ldots,J\}$ (e.g.,
cancer or death without cancer), and $p$-dimensional covariate vector
$X_{ik}$ for twin pair $i=1,\ldots,n$ and individual $k=1,2$.  We will
assume that the $n$ pairs $\{T_i,C_i,\cause_i,X_i\} =
\{\{T_{i1},T_{i2}\},\{C_{i1},C_{i2}\},\{\cause_{i1},\cause_{i2}\},\{X_{i1},X_{i2}\}\}$
are iid. Due to the right-censoring, we only observe $\otime_{ik} =
T_{ik}\wedge C_{ik}$ and $\ocause_{ik}=\cause_{ik}\Delta_{ik}$, with
the indicator for $T_{ik}$ denoting an actual event time $\Delta_{ik}
= I(T_{ik}\leq C_{ik})$. We will perceive the data as generated by the
model described by the diagram in Figure~\ref{fig:comprisk}, where
every subject starts in the alive state, and then moves to either of
the two states prostate cancer of death with certain intensities
evolving over time.  Note that in our application, we are not aiming to
make inference on the transition from prostate cancer to death.

In the univariate setting, the transition may be characterized by the
cumulative incidence functions
\begin{align*}
  F_1(t) = \pr(T\leq t,\cause=1),
\end{align*}
which may be estimated by the Aalen-Johansen estimator
\citep{aale:joha:1978,andersencounting} and also generalized to the
regression setting as in \cite{scheikebinomialregr08}. The
bivariate case is more complex but the concordance function
\begin{align*}
  \mathcal{C}(t) = \pr(T_1\leq t,T_2\leq t, \cause_1=1, \cause_2=1),
\end{align*}
may be estimated as described in \cite{scheike13:concordance}. Here we
will only consider a fixed time $\tau$ 
% (e.g.,full lifespan
% $\tau=\infty$)
and characterize the joint probability
\begin{align}\label{eq:tau}
 \pr(T_1\leq\tau, T_2\leq\tau, \cause_1=1,\cause_2=1),
\end{align}
which we will assume can be modelled by a random effect
structure as in \eqref{eq:liabilityprob}
\begin{align}\label{eq:liabilitycomprisk}
  \pr(T_{ij}\leq\tau, \cause_{ij}=1 | \latent_{ij},X_{ij}) = \Phi(\beta^\transpose
  X_{ij} + \latent_{ij}).
\end{align}
We will use age as our time-scale, and assuming that everyone were
followed until time $\tau$ this simply corresponds to a standard
liability model where twins are classified as having cancer or not
before time $\tau$, in which case the standard MLE approach of
\eqref{eq:liabilityprob} would be consistent. In practice, a large
fraction of the twins may not have reached the age $\tau$ at the end
of follow-up, and other techniques must be applied.

\subsection{Consistent Estimating Equations}

In this section we will introduce inverse probability weighting to
correct for the right censoring. The intuition for this procedure is
that the observations that have a higher probability of being censored
are under-represented and should therefore count more in the analysis.
These techniques can be traced back to the Horwitz-Thompson estimator
applied in the survey-statistics field \citep{horvitzthompson1952} and
later with many applications in other fields of statistics for dealing
with coarsened data including survival analysis
\citep{rotnitzkyrobbins1995,robins92} and competing risks
\citep{fine_and_gray_competing_risk_1999}. We refer to
\cite{tsiatis2006semiparametric} for a modern and accessible treatment
of the subject in both the parametric and semi-parametric
setting. Here we are interested in estimating dependence between
paired observations which in general complicates the analysis, due to
the need of consistent estimates of the bivariate censoring
probabilities. We will show how the complexity may be reduced
dramatically by exploiting how data is collected in registry studies.

The full-data score equation we obtain from the model
\eqref{eq:liabilitycomprisk} parametrised by $\theta$ (including both
$\beta$ and the parameters of the random effects), when all subjects
are followed until time $\tau$, will be denoted
\begin{align}\label{eq:u0}
  \mathcal{U}_0(\theta;X,\otime,\ocause) = \sum_{i=1}^n
  \mathcal{U}_{0i}(\theta; X_{i},\otime_i,\ocause_{i}),
\end{align}
where $\mathcal{U}_{0i}(\cdot;X_{i},\otime_i,\ocause_{i})$ is the
derivative of the log-likelihood term for a bivariate Probit model
\citep{ashford70probit} for the event
$(\epsilon_{ij}=1,\otime_{ij}\leq\tau)$ of the $i$th twin-pair. A nice
property of the Probit random effects model is that the marginal
distribution obtained by integrating over the normal distributed
random effects is also a multivariate Probit model, and the derivative
of the log-likelihood with respect to the parameter vector may in turn
be written as a linear combination of bivariate cumulative normal
distribution functions.  The general derivation may be found in
\citep{holst:binarylatent}, and the integration problem related to
evaluating the bivariate cumulative distribution functions can be
dealt with as described in
\citep{genz92:_numer_comput_of_multiv_normal_probab}. In principle, the
same procedure could be applied to higher-dimensional problems thus
allowing us to generalize the modelling framework to larger pedigrees.

We will describe the censoring distribution by its survival function
\begin{align}\label{eq:jointcens}
  \Gc(t_1,t_2; Z_{i}) = \pr(C_{i1}>t_1,C_{i2}>t_2\mid Z_{i}),
\end{align}
given covariates $Z_{i}$ observed for all twin-pairs
$i=1,\ldots,n$, and we will assume that the failure times are
independent of the censoring times given these covariates.

Furthermore, we will assume that we have a correct model for the censoring
mechanism with estimate $\hatGc$. We then define the IPCW-adjusted
estimating equation via the new score 
\begin{align}\label{eq:ipcw1}
  \mathcal{U}(\theta; X,Z,\otime,\ocause) = \sum_{i=1}^n
  \mathcal{U}_i(\theta; X_i,Z_i,\otime_i,\ocause_i) = \sum_{i=1}^n
  \frac{\Delta_{i1}\Delta_{i2}}{\hatGc(\otime_{i1},\otime_{i2};
    Z_{i})}\mathcal{U}_{0i}(\theta; X_i,\otime_i,\ocause_i).
\end{align}

The censoring mechanism \eqref{eq:jointcens} may be modelled using
frailty models, but in the case where data arises from a twin registry,
censoring will typically be administrative and hence twins are
censored at the same time. In this case
\begin{align}\label{eq:minGc}
  \Gc(t_1,t_2\mid Z_{i}) = \pr(C_{i}> t_1\vee t_2 \mid Z_{i}) =
  \Gc(t_1\mid Z_{i})\wedge \Gc(t_2\mid Z_{i}).
\end{align}
Therefore, the problem of identifying the bivariate censoring distribution
is simplified to just estimating the marginal
censoring distributions. 
%As noted by \cite{scheike13:concordance} 

Consistency of the parameter estimates relies on a correctly
specified model for the censoring mechanism \eqref{eq:minGc}, which
would suggest a quite rich semi-parametric model for the marginal
censoring distributions. 
%% In large register studies this will
%%typically not be a limiting assumption.
%in the sense that slightly over-fitting will be
%better than under-fitting the model. 
However, a computational limitation of the semi-parametric
approach is, that the calculation of asymptotic standard errors (from
the estimated influence functions as described below) is quite
computational intensive in the order $\mathcal{O}(n K)$ where $K$ is
the number of event times and $n$ the number of subjects. In large
registry studies a sufficiently flexible parametric survival model may
therefore be preferable.  We note that asymptotic double-robustness
could be obtained by adding an augmentation term to the estimating
equation \citep{tsiatis2006semiparametric} requiring just on of the
two models to be correct to obtain consistency. In the following, we
will, however, assume that $\Gc$ lies within a parametric family and
let $\widehat{\gamma}$ be a consistent estimator such that
$\hatGc(\cdot;z) = \hatGc(\cdot;z,\widehat{\gamma})$.

\begin{thm}\label{thm1}
  Let $\{T_i,C_i,\cause_i,X_i,Z_i\}$ be iid and $\widehat{\gamma}$
  a consistent regular asymptotic linear estimator for the
  parametric censoring distribution. Denote the right-hand-side
  terms of \eqref{eq:ipcw1} as $\mathcal{U}_i(\theta_0,\gamma_0)$.
  Under the following regularity conditions
  \begin{enumerate}
  \item\label{cond:deriv} In a neighbourhood of $(\theta_0^\transpose,\gamma_0^\transpose)^\transpose$ the function
    $\mathcal{U}_i$ is twice continuous differentiable with 
    $\E(-\partial\mathcal{U}(\theta_0,\gamma_0)/\partial\theta)$ being
    positive-definite.
  \item \label{cond:indep} The censoring times $(C_{1i},C_{2i})$ are conditionally
    independent of $(T_{1i},T_{2i},\epsilon_{1i},\epsilon_{2i})$ implying
    $G_c(t_1-,t_2-; z) = \E(\Delta_{1i}\Delta_{2i} \mid
    T_{1i}=t_1,T_{2i}=t_2,Z_{i}=z)$.
  \item\label{cond:atrisk} $\pr(T_{1i}>\tau,T_{2i}>\tau)>0$
  \item\label{cond:bounded} The covariates $X_{i}, Z_{i}$ are bounded.
  \item\label{cond:positivity} $G_c(t_1,t_2; z)>0$ with probability 1 for $t_1,t_2\in
    [0,\tau]$.
  \end{enumerate}  
  the estimator $\widehat{\theta}$ obtained as the root of
  \eqref{eq:ipcw1} is consistent and asymptotically normal.
\end{thm}

Consistency follows from condition \ref{cond:deriv}-\ref{cond:atrisk}
by noting that for any term on the right-hand-side of \eqref{eq:ipcw1},
we obtain for known censoring distribution:
\begin{align*}
  \E[\mathcal{U}_i(\theta;X_i,Z_i,\otime_i,\ocause_i)] &=
  \E\{\E[\mathcal{U}_i(\theta;X_i,Z_i,\otime_i,\ocause_i)\mid
  X_i,Z_i,\otime_i,\ocause_i]\} \\
  &=  \E\{\E(\Delta_{i1},\Delta_{i2}\mid Z_i,\otime_i,\ocause_i)
  G_c(\otime_{i1},\otime_{i2}\mid Z_i)^{-1}\mathcal{U}_{0i}(\theta;
  X_i,\otime_i,\ocause_i)\} \\
  &= \E[\mathcal{U}_{0i}(\theta; X_i,\otime_i,\ocause_i)] = 0,
\end{align*}
where we actively assumed
% used conditional independence between failure times and
% censoring times and
consistency of both the models \eqref{eq:liabilitycomprisk} and
\eqref{eq:jointcens}. Note that the positive probability of being
at risk is fulfilled when the support of the censoring times lies
within the support of $T_{1i}$ and $T_{2i}$.
% From condition
% \ref{cond:bounded} follows that the empirical
% averages of the derivatives of the score converges to their respective
% expectations allowing us to make Taylor expansion in \eqref{eq:IF3}
% below.  
We emphasize that
a key regularity condition here is that of positivity
\eqref{cond:positivity}, namely that the probability of any twin-pair
being uncensored is strictly larger than zero. In practice, the
probabilities should be sufficiently large to avoid instability of the
estimating equation in smaller sample sizes.
% In the following we will denote the marginal censoring distribution
% \begin{align*}
%   \Gc(t; X_{ik}) = \pr(C_{ik}>t\mid X_{ki})
% \end{align*}
% and the corresponding joint distribution using the same notation

We now sketch the calculation of the asymptotic standard errors of
the estimator.  The estimator for $\widehat{\gamma}$ will typically be
a GEE-type $m$-estimator since we will use both twins to estimate the
marginal censoring distribution. This implies asymptotic linearity:
\begin{align*}
  \sqrt{n}(\widehat{\gamma}-\gamma_0) = n^{-1/2}\sum_{i=1}^n
  \influ_1(\gamma_0; Z_i,\otime_i,\ocause_i) + o_p(1),
\end{align*}
where $\influ_1$ is the influence function of the estimator
\citep{stefanski_boos_2002_m-estimator}.

Let $\widehat{\theta}(\widehat{\gamma})$ be the two-stage estimator
obtained by finding the root of  \eqref{eq:ipcw1} with the plugin-estimate
of the censoring distribution via $\widehat{\gamma}$.  The conditions
of Theorem \ref{thm1} implies that the empirical averages of the derivatives of
the score converges to their corresponding expectations, and a 
Taylor expansion of \eqref{eq:ipcw1} around the true parameters
$\theta_0$ and $\gamma_0$, shows that
\begin{align}
  \begin{split}\label{eq:IF3}
    \sqrt{n}(\widehat{\theta}(\widehat{\gamma}) - \theta_0)
    &= n^{-1/2}\sum_{i=1}^n \influ_2(\theta_0; X_i,Z_i,\otime_i,\ocause_i)  \\
    &\hspace*{-6ex}+ n^{-1/2}\,\E[\frac{\partial}{\partial\theta}\mathcal{U}_i(\theta_0,\gamma_0)]^{-1}
    \E[\frac{\partial}{\partial\gamma} \mathcal{U}_i(\theta_0,\gamma_0)]
    \sum_{i=1}^n\influ_1(\gamma_0; Z_i,\otime_i,\ocause_i) + o_p(1) \\
    &= 
    n^{-1/2}\sum_{i=1}^n \influ_3(\theta_0; X_i,Z_i,\otime_i,\ocause_i,) + o_p(1),
  \end{split}
\end{align}
where the first term corresponds to the iid decomposition for known
censoring distribution
\begin{align}\label{eq:IF2}
  \sqrt{n}(\widehat{\theta}(\gamma_0)-\theta_0) = n^{-1/2}\sum_{i=1}^n
  \influ_2(\theta_0; X_i,\otime_i,\ocause_i) + o_p(1).
\end{align}
The influence functions may be estimated from the bi-products of the
Newton-Raphson optimization, as the matrix product of the derivative of
the score times the score itself. We refer to
\cite{holst:binarylatent} for expressions for the relevant terms of
$\influ_2$, which are implemented in the \texttt{mets} \texttt{R}-package
\citep{metspackage}.

It follows from \eqref{eq:IF3} that the two-stage estimator is
asymptotically normal and the asymptotic variance of \eqref{eq:IF3}
can be estimated by plugging in the parameter estimates
\begin{align*}
  \frac{1}{n}\sum_{i=1}^n
  \influ_3(\widehat{\theta},\widehat{\gamma}; X_i,Z_i,\otime_i,\ocause_i)^{\otimes 2}.
\end{align*}

Similar results can be shown in the general case where $\hatGc$ is
an asymptotically linear consistent estimator of the censoring
distribution, such that
  \begin{align*}
    \sqrt{n}\left[\hatGc(t_1,t_2;Z)-\Gc(t_1,t_2;Z)\right] =
    n^{-1/2}\sum_{i=1}^n\influ_{\Gc}(t_1,t_2,Z;
    Z_i,\otime_{i},\ocause_{i})
      + o_p(1),
  \end{align*}
where the iid terms $\influ_{\Gc}$ are the influence functions. 
For the choice of a Cox-regression, the proof of the consistency and
asymptotic normality of the IPCW estimator follows along the lines of
\cite{scheikebinomialregr08} or \cite{linmedicalcost2000}.

In the case of a Kaplan-Meier estimator the linear expansion above
follows from \cite{gill-thesis}; see also Section IV.3.2 of
\cite{andersencounting}. In the case of a Cox model the linear
expansion is a consequence of the results in Section VII.2.2 and
VII.2.3 of \cite{andersencounting}. Specific technical assumptions are
also given there.  Here, the focus is on the use of parametric models
due to the computational advantages.

% We take an approach that has the same aim but considers the
% time-aspect and corrects for censorings and thus leads to
% interpretable estimates.  We take censoring into account by inverse
% probability censoring weighting (IPCW) techniques.
% \begin{itemize}
% \item IPCW: Aalen or KM. Same censoring in pairs. Figure explaining
%   the ipcw idea applied to same censored twin data.
% \item Estimation: The method allows for twin singletons. Previously the cohort of only intact pairs are in use.
% \item Implementation.
% \item Features: Concordance, Covariates, Efficient (since
%   probit-model).
% \end{itemize}

\subsection{Model Selection and Testing}\label{sec:modelselect}

The main hypothesis in most applications of the Liability Threshold
model will be to \textit{a)} Test for a genetic component \textit{b)}
Quantify this effect. The first problem should generally not be
examined in the polygenic model to avoid in part the many genetic
model assumptions and in part the difficulties of testing parameters
on the boundary of the parameter space.  A reasonable modelling
approach is generally to initially estimate a more flexible model,
where we instead of a random effects model estimate the parameters of
a bivariate Probit model
\begin{align}\label{eq:biprobit}
  \pr(T_1\leq\tau ,T_2\leq\tau\,\epsilon_1=1,\epsilon_2=1, \mid X_1,X_2) = 
  \Phi_{\rho_{\text{zyg}}}(\beta_{\text{zyg}}^\transpose X_1,\beta_{\text{zyg}}^\transpose X_2^\transpose),
\end{align}
where $\Phi_{\rho_{\text{zyg}}}$ is the bivariate normal CDF with mean
0 and variance given by a correlation matrix with correlation
coefficient $_{\rho_{\text{zyg}}}$ depending on zygosity. A test for
identical marginals should be done as a first step, i.e., testing if
$\beta_{\text{MZ}}=\beta_{\text{DZ}}$. Next, a formal test for the
presence of a genetic component can be obtained by testing the null
hypothesis of identical tetrachoric correlations in MZ and DZ twins
$\rho_{\text{MZ}}=\rho_{\text{DZ}}$. Estimates on the risk scale such
as concordance rates are also preferably calculated in this model.
Note that while the test for genetic influence still requires
assumption of same environmental effects in MZ and DZ twins, the many
genetic assumptions of the polygenic model, e.g., linkage
equilibrium and that a subset of ACDE fits the data, are no longer
necessary.

With evidence of a genetic component, the next step should be to
quantify the possible genetic and environmental effects based on the
IPCW adjusted Liability Threshold model \eqref{eq:liabilitycomprisk}.
In population genetics it is common to compare different models using
information criteria such as the AIC \citep{Akaike73}.  In general, the
derivation of these measures relies on inference being done within
a maximum likelihood framework, and are no longer generally valid in an
estimating equation framework. The Quasi-AIC (QIC) has been suggested
\citep{QICpan01} in the GEE framework. However, in the case of
\eqref{eq:liabilitycomprisk} the estimating equation corresponds to the
weighted score-function of the complete-data likelihood  $\sum_{i=1}^n
\log L_i(\theta; X_i,\otime_i,\ocause_i)$ from which \eqref{eq:u0} is obtained.
It follows that 
\begin{align*}
  \E\left[\frac{\Delta_{i1}\Delta_{i2}}{\Gc(T_{i1},T_{i2};
  Z_{i})}\log L_i(\theta; X_i,\otime_i,\ocause_i)\right] = \E(\log
L_i(\theta; X_i,\otime_i,\ocause_i)),
\end{align*}
and hence the weighted AIC
\begin{align*}
  \wAIC =
  2\sum_{i=1}^n\frac{\Delta_{i1}\Delta_{i2}}{\hatGc(T_{i1},T_{i2}; Z_i)}\log
  L_i(\theta; X_i,\otime_i,\ocause_i) - 2P,
\end{align*}
where $P$ is the number of parameters in $\theta$, will also provide an
approximation of the relative entropy between the estimated model and
the true data generating model, and may therefore serve as a model
selection tool.

% Kullback-Leibler divergence, relative entropy... As for the
% consistency of the parameter estimates relies on a correctly specified
% model for the censoring mechanism the same applies here for the valid
% interpretation of the weighted AIC.

%}}} Comp.risk & IPCW

%{{{ Simulation

\section{Simulation study}\label{sec:sim}

\begin{figure}[htbp]
  \hfuzz=12pt
  \centering
    \mbox{
      \includegraphics[width=0.5\textwidth]{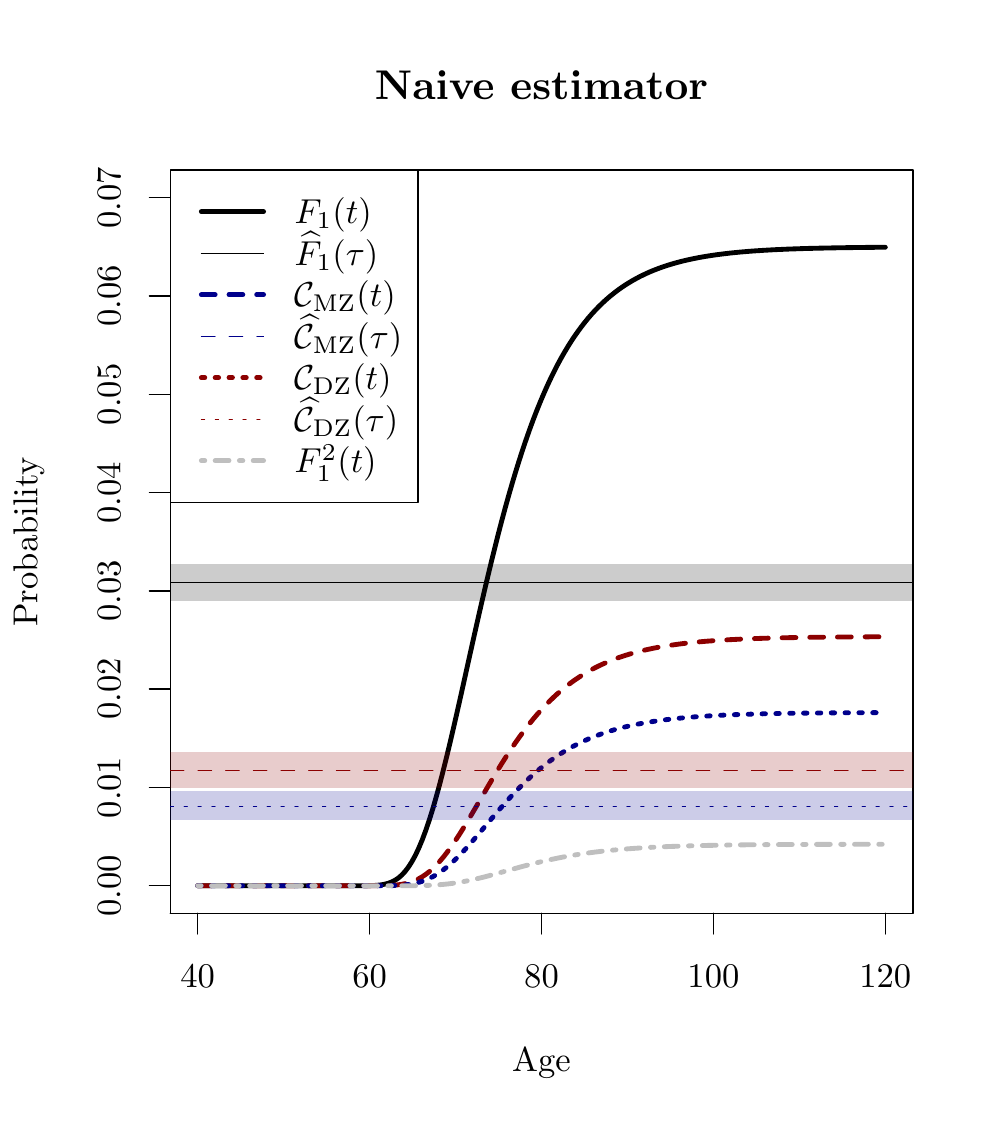}
      \includegraphics[width=0.5\textwidth]{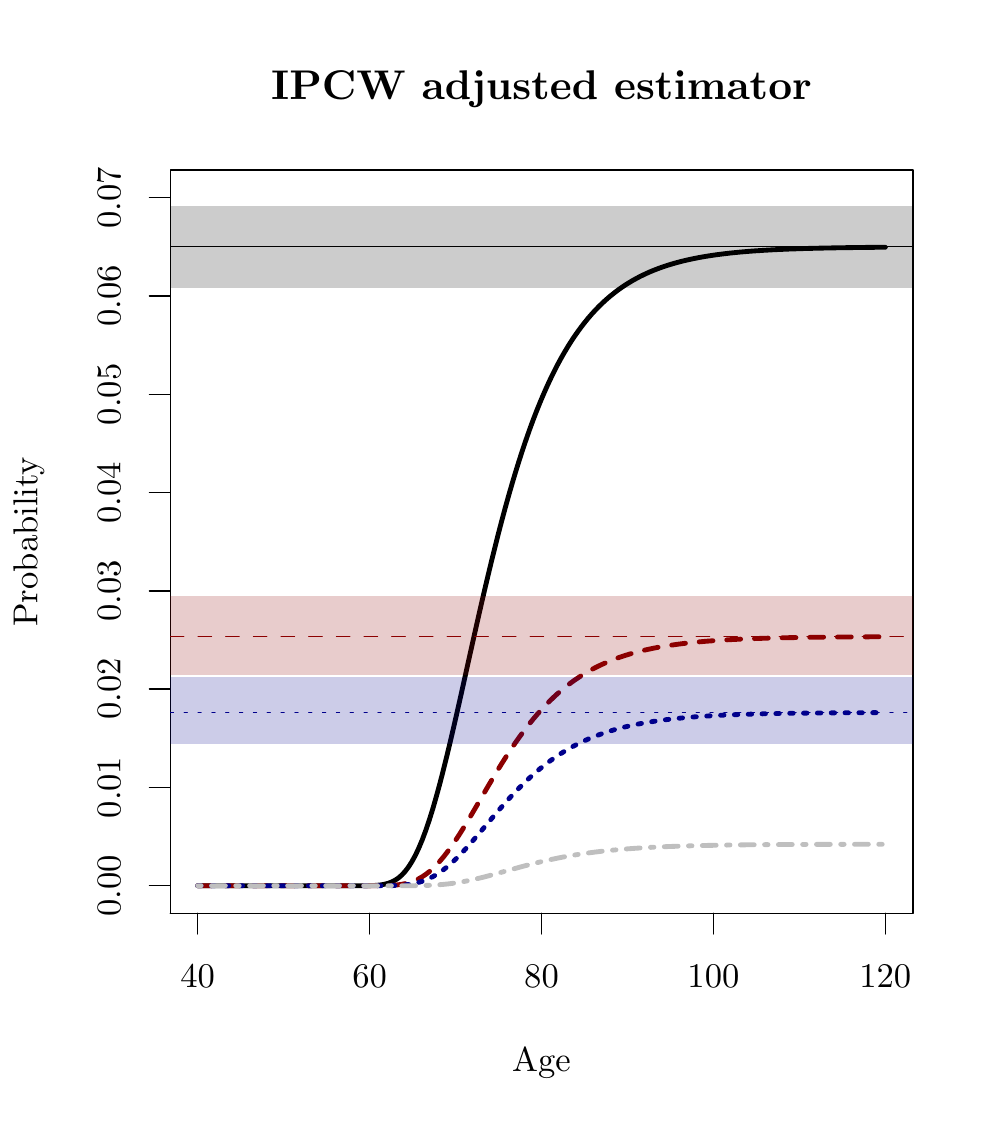}
    }
    \caption{Simulated cumulative incidence and concordance function
      with $\sigma_A^2=\sigma_C^2=\sigma_E^2=\tfrac{1}{3}$. Thick
      lines shows true cumulative incidence for cancer ($F_1$,
      benchmark for perfect dependence), MZ concordance function
      ($\mathcal{C}_{\text{MZ}}$), DZ concordance function
      ($\mathcal{C}_{\text{DZ}}$), and the squared cumulative
      incidence ($F_1^2$, benchmark for independence). The thin
      horizontal lines shows the mean estimates and 2.5\% and 97.5\%
      quantiles of 1,000 replications with 20,000 twin pairs and 59\%
      censoring, for the naive estimator ignoring censoring (left
      panel) and the IPCW adjusted estimator (right
      panel).\label{fig:simflat}}
\end{figure}

We set up a simulation study to examine the properties of our proposed
estimator in a realistic setup. The cumulative incidence function for
cancer conditional on a random effect $\eta_1$, was chosen as
\begin{align*}
  F_1(t\mid \eta_1)= \pr(T\leq t,\epsilon=1\mid \eta_1) =
  \Phi_{\sigma_{E_1^2}}(\alpha(t)+\Phi^{-1}(p_1)+\eta_1),
\end{align*}
with $\alpha(t) = -\exp(10-0.15t)), \ p_1=0.065$. The inverse link-function
$\Phi_{\sigma_{E_1}^2}$ was chosen as a normal CDF with variance
$\sigma_{E_1^2} = 1-\var(\eta_1)$.  This parametrisation leads to a
marginal CIF resembling the distribution observed in the real data
described in Section~\ref{sec:application}, with a marginal lifetime
prevalence of 0.065 (see Figure~\ref{fig:simflat}). The type of cause
(cancer or death without cancer) were simulated from a
Bernoulli-distribution with probability $F_1(\infty) =
\Phi_{\sigma_{E^2_1}}(\eta_1+\Phi^{-1}(0.065))$, and the event times
drawn from $\pr(T\leq t \mid \eta_k,\epsilon=k)$ which for the
competing risk of death was chosen as the distribution
$\Phi_{\sigma_{E_2^2}}(0.1(t-85) + \eta_2)$, again with a marginal
resembling what was observed in the real data example.  The random
effect structure $\eta_1$ was chosen as an ACE-model with the
C-component shared across the two competing risks, and with $\eta_2$
only consisting of this shared environmental effect $\eta_2 =
\eta^{C}$.  Independent censoring was simulated from a Weibull
distribution with cumulative hazard $\Lambda_0(t) = (\lambda t)^\nu$,
with scale parameter fixed at $\log(\lambda)=-4.5$, and the parameters
were estimated using a marginal model with working independence
structure.

We simulated 10,000 MZ and 10,000 DZ twin pairs from the above model
under three different ACE structures
$(\sigma_A^2,\sigma_C^2,\sigma_C^2)\in
\{(\tfrac{1}{3},\tfrac{1}{3},\tfrac{1}{3}),
(\tfrac{1}{2},\tfrac{1}{4},\tfrac{1}{4}),
(\tfrac{3}{5},\tfrac{1}{5},\tfrac{1}{5})\}$, and with varying degree
of censoring $\log(\nu)\in\{0.5,2\}$ corresponding to roughly 59\% and
48\% right-censoring. In each scenario the naive estimator ignoring
censoring was compared to the IPCW-adjusted estimators based on a
parametric marginal Weibull model, with standard errors based on the
correct influence functions \eqref{eq:IF3} (Weibull${}_2$) and 
standard errors based on the influence function \eqref{eq:IF2} without
adjusting for the uncertainty in the weights (Weibull${}_1$), and an
IPCW-adjusted estimator based on the Kaplan-Meier estimator (KM).

\begin{sidewaystable}
\centering
\footnotesize
%%\scriptsize
\begin{tabular}{llcccccccccccc}
\toprule
&& \multicolumn{2}{c}{$F_1$} & \multicolumn{2}{c}{$\mathcal{C}_{MZ}$} & \multicolumn{2}{c}{$\mathcal{C}_{DZ}$} & \multicolumn{2}{c}{$\sigma_A^2$} & \multicolumn{2}{c}{$\sigma_C^2$} & \multicolumn{2}{c}{$\sigma_E^2$}
\\ \cmidrule(lr){3-4} \cmidrule(lr){5-6} \cmidrule(lr){7-8} \cmidrule(lr){9-10} \cmidrule(lr){11-12} \cmidrule(lr){13-14}
 &   & Av. & Cv. & Av. & Cv. & Av. & Cv. & Av. & Cv. & Av. & Cv. & Av. & Cv. \\
\midrule   & True & 0.065 &     & 0.025 &     & 0.018 &     & 0.333 &
& 0.333 &     & 0.333 &     \\ 
\cmidrule(lr){2-14}
\multirow{14}{*}{\parbox{5em}{$\nu=1.6$\\ 59\% cens.}} & \cellcolor{tableShade}Naive & \cellcolor{tableShade}0.031 & \cellcolor{tableShade}0.000 & \cellcolor{tableShade}0.012 & \cellcolor{tableShade}0.000 & \cellcolor{tableShade}0.008 & \cellcolor{tableShade}0.000 & \cellcolor{tableShade}0.280 & \cellcolor{tableShade}0.888 & \cellcolor{tableShade}0.452 & \cellcolor{tableShade}0.559 & \cellcolor{tableShade}0.267 & \cellcolor{tableShade}0.212\\ 
 & \cellcolor{white}Weibull${}_1$ & \cellcolor{white}0.065 & \cellcolor{white}0.948 & \cellcolor{white}0.025 & \cellcolor{white}0.944 & \cellcolor{white}0.018 & \cellcolor{white}0.956 & \cellcolor{white}0.335 & \cellcolor{white}0.957 & \cellcolor{white}0.331 & \cellcolor{white}0.956 & \cellcolor{white}0.334 & \cellcolor{white}0.940\\ 
 & \cellcolor{tableShade}Weibull${}_2$ & \cellcolor{tableShade}0.065 & \cellcolor{tableShade}0.948 & \cellcolor{tableShade}0.025 & \cellcolor{tableShade}0.944 & \cellcolor{tableShade}0.018 & \cellcolor{tableShade}0.957 & \cellcolor{tableShade}0.335 & \cellcolor{tableShade}0.957 & \cellcolor{tableShade}0.331 & \cellcolor{tableShade}0.956 & \cellcolor{tableShade}0.334 & \cellcolor{tableShade}0.940\\ 
 & \cellcolor{white}KM & \cellcolor{white}0.065 &
 \cellcolor{white}0.948 & \cellcolor{white}0.025 &
 \cellcolor{white}0.944 & \cellcolor{white}0.018 &
 \cellcolor{white}0.955 & \cellcolor{white}0.335 &
 \cellcolor{white}0.957 & \cellcolor{white}0.331 &
 \cellcolor{white}0.955 & \cellcolor{white}0.334 &
 \cellcolor{white}0.940\\ 
\cmidrule(lr){2-14}
 & True & 0.065 &     & 0.030 &     & 0.018 &     & 0.500 &     &
 0.250 &     & 0.250 &     \\ 
\cmidrule(lr){2-14}
 & \cellcolor{tableShade}Naive & \cellcolor{tableShade}0.031 & \cellcolor{tableShade}0.000 & \cellcolor{tableShade}0.014 & \cellcolor{tableShade}0.000 & \cellcolor{tableShade}0.008 & \cellcolor{tableShade}0.000 & \cellcolor{tableShade}0.414 & \cellcolor{tableShade}0.769 & \cellcolor{tableShade}0.386 & \cellcolor{tableShade}0.453 & \cellcolor{tableShade}0.200 & \cellcolor{tableShade}0.273\\ 
 & \cellcolor{white}Weibull${}_1$ & \cellcolor{white}0.065 & \cellcolor{white}0.952 & \cellcolor{white}0.030 & \cellcolor{white}0.952 & \cellcolor{white}0.018 & \cellcolor{white}0.953 & \cellcolor{white}0.498 & \cellcolor{white}0.956 & \cellcolor{white}0.250 & \cellcolor{white}0.956 & \cellcolor{white}0.252 & \cellcolor{white}0.946\\ 
 & \cellcolor{tableShade}Weibull${}_2$ & \cellcolor{tableShade}0.065 & \cellcolor{tableShade}0.952 & \cellcolor{tableShade}0.030 & \cellcolor{tableShade}0.952 & \cellcolor{tableShade}0.018 & \cellcolor{tableShade}0.953 & \cellcolor{tableShade}0.498 & \cellcolor{tableShade}0.956 & \cellcolor{tableShade}0.250 & \cellcolor{tableShade}0.956 & \cellcolor{tableShade}0.252 & \cellcolor{tableShade}0.946\\ 
 & \cellcolor{white}KM & \cellcolor{white}0.065 &
 \cellcolor{white}0.954 & \cellcolor{white}0.030 &
 \cellcolor{white}0.954 & \cellcolor{white}0.018 &
 \cellcolor{white}0.954 & \cellcolor{white}0.498 &
 \cellcolor{white}0.957 & \cellcolor{white}0.250 &
 \cellcolor{white}0.955 & \cellcolor{white}0.252 &
 \cellcolor{white}0.945\\ 
\cmidrule(lr){2-14}
  & True & 0.065 &     & 0.034 &     & 0.018 &     & 0.600 &     &
  0.200 &     & 0.200 &     \\ 
\cmidrule(lr){2-14}
 & \cellcolor{tableShade}Naive & \cellcolor{tableShade}0.031 & \cellcolor{tableShade}0.000 & \cellcolor{tableShade}0.016 & \cellcolor{tableShade}0.000 & \cellcolor{tableShade}0.008 & \cellcolor{tableShade}0.000 & \cellcolor{tableShade}0.491 & \cellcolor{tableShade}0.636 & \cellcolor{tableShade}0.349 & \cellcolor{tableShade}0.365 & \cellcolor{tableShade}0.160 & \cellcolor{tableShade}0.327\\ 
 & \cellcolor{white}Weibull${}_1$ & \cellcolor{white}0.065 & \cellcolor{white}0.946 & \cellcolor{white}0.034 & \cellcolor{white}0.952 & \cellcolor{white}0.018 & \cellcolor{white}0.939 & \cellcolor{white}0.593 & \cellcolor{white}0.950 & \cellcolor{white}0.204 & \cellcolor{white}0.946 & \cellcolor{white}0.203 & \cellcolor{white}0.950\\ 
 & \cellcolor{tableShade}Weibull${}_2$ & \cellcolor{tableShade}0.065 & \cellcolor{tableShade}0.946 & \cellcolor{tableShade}0.034 & \cellcolor{tableShade}0.953 & \cellcolor{tableShade}0.018 & \cellcolor{tableShade}0.942 & \cellcolor{tableShade}0.593 & \cellcolor{tableShade}0.954 & \cellcolor{tableShade}0.204 & \cellcolor{tableShade}0.950 & \cellcolor{tableShade}0.203 & \cellcolor{tableShade}0.951\\ 
 & \cellcolor{white}KM & \cellcolor{white}0.065 &
 \cellcolor{white}0.945 & \cellcolor{white}0.034 &
 \cellcolor{white}0.952 & \cellcolor{white}0.018 &
 \cellcolor{white}0.939 & \cellcolor{white}0.593 &
 \cellcolor{white}0.951 & \cellcolor{white}0.204 &
 \cellcolor{white}0.948 & \cellcolor{white}0.203 &
 \cellcolor{white}0.952\\ 
\midrule
 & True & 0.065 &     & 0.025 &     & 0.018 &     & 0.333 &     &
 0.333 &     & 0.333 &     \\ 
\midrule
\multirow{14}{*}{\parbox{5em}{$\nu=7.4$\\48\% cens.}} & \cellcolor{tableShade}Naive & \cellcolor{tableShade}0.048 & \cellcolor{tableShade}0.000 & \cellcolor{tableShade}0.018 & \cellcolor{tableShade}0.000 & \cellcolor{tableShade}0.012 & \cellcolor{tableShade}0.000 & \cellcolor{tableShade}0.318 & \cellcolor{tableShade}0.951 & \cellcolor{tableShade}0.366 & \cellcolor{tableShade}0.907 & \cellcolor{tableShade}0.315 & \cellcolor{tableShade}0.850\\ 
 & \cellcolor{white}Weibull${}_1$ & \cellcolor{white}0.065 & \cellcolor{white}0.955 & \cellcolor{white}0.025 & \cellcolor{white}0.948 & \cellcolor{white}0.018 & \cellcolor{white}0.951 & \cellcolor{white}0.332 & \cellcolor{white}0.953 & \cellcolor{white}0.333 & \cellcolor{white}0.955 & \cellcolor{white}0.334 & \cellcolor{white}0.949\\ 
 & \cellcolor{tableShade}Weibull${}_2$ & \cellcolor{tableShade}0.065 & \cellcolor{tableShade}0.955 & \cellcolor{tableShade}0.025 & \cellcolor{tableShade}0.948 & \cellcolor{tableShade}0.018 & \cellcolor{tableShade}0.953 & \cellcolor{tableShade}0.332 & \cellcolor{tableShade}0.955 & \cellcolor{tableShade}0.333 & \cellcolor{tableShade}0.956 & \cellcolor{tableShade}0.334 & \cellcolor{tableShade}0.950\\ 
 & \cellcolor{white}KM & \cellcolor{white}0.065 & \cellcolor{white}0.956 & \cellcolor{white}0.025 & \cellcolor{white}0.950 & \cellcolor{white}0.018 & \cellcolor{white}0.955 & \cellcolor{white}0.333 & \cellcolor{white}0.954 & \cellcolor{white}0.332 & \cellcolor{white}0.954 & \cellcolor{white}0.335 & \cellcolor{white}0.953\\ 
\cmidrule(lr){2-14}
 & True & 0.065 &     & 0.030 &     & 0.018 &     & 0.500 &     &
 0.250 &     & 0.250 &     \\ 
\cmidrule(lr){2-14}
 & \cellcolor{tableShade}Naive & \cellcolor{tableShade}0.048 & \cellcolor{tableShade}0.000 & \cellcolor{tableShade}0.021 & \cellcolor{tableShade}0.000 & \cellcolor{tableShade}0.012 & \cellcolor{tableShade}0.001 & \cellcolor{tableShade}0.477 & \cellcolor{tableShade}0.936 & \cellcolor{tableShade}0.287 & \cellcolor{tableShade}0.896 & \cellcolor{tableShade}0.236 & \cellcolor{tableShade}0.865\\ 
 & \cellcolor{white}Weibull${}_1$ & \cellcolor{white}0.065 & \cellcolor{white}0.946 & \cellcolor{white}0.030 & \cellcolor{white}0.965 & \cellcolor{white}0.018 & \cellcolor{white}0.938 & \cellcolor{white}0.496 & \cellcolor{white}0.952 & \cellcolor{white}0.252 & \cellcolor{white}0.945 & \cellcolor{white}0.252 & \cellcolor{white}0.950\\ 
 & \cellcolor{tableShade}Weibull${}_2$ & \cellcolor{tableShade}0.065 & \cellcolor{tableShade}0.946 & \cellcolor{tableShade}0.030 & \cellcolor{tableShade}0.966 & \cellcolor{tableShade}0.018 & \cellcolor{tableShade}0.941 & \cellcolor{tableShade}0.496 & \cellcolor{tableShade}0.958 & \cellcolor{tableShade}0.252 & \cellcolor{tableShade}0.952 & \cellcolor{tableShade}0.252 & \cellcolor{tableShade}0.952\\ 
 & \cellcolor{white}KM & \cellcolor{white}0.065 & \cellcolor{white}0.958 & \cellcolor{white}0.030 & \cellcolor{white}0.964 & \cellcolor{white}0.018 & \cellcolor{white}0.942 & \cellcolor{white}0.498 & \cellcolor{white}0.957 & \cellcolor{white}0.251 & \cellcolor{white}0.952 & \cellcolor{white}0.251 & \cellcolor{white}0.957\\ 
\cmidrule(lr){2-14}
 & True & 0.065 &     & 0.034 &     & 0.018 &     & 0.600 &     &
 0.200 &     & 0.200 &     \\ 
\cmidrule(lr){2-14}
 & \cellcolor{tableShade}Naive & \cellcolor{tableShade}0.048 & \cellcolor{tableShade}0.000 & \cellcolor{tableShade}0.024 & \cellcolor{tableShade}0.000 & \cellcolor{tableShade}0.012 & \cellcolor{tableShade}0.003 & \cellcolor{tableShade}0.570 & \cellcolor{tableShade}0.918 & \cellcolor{tableShade}0.240 & \cellcolor{tableShade}0.877 & \cellcolor{tableShade}0.189 & \cellcolor{tableShade}0.871\\ 
 & \cellcolor{white}Weibull${}_1$ & \cellcolor{white}0.065 & \cellcolor{white}0.952 & \cellcolor{white}0.034 & \cellcolor{white}0.940 & \cellcolor{white}0.018 & \cellcolor{white}0.940 & \cellcolor{white}0.598 & \cellcolor{white}0.922 & \cellcolor{white}0.201 & \cellcolor{white}0.924 & \cellcolor{white}0.201 & \cellcolor{white}0.939\\ 
 & \cellcolor{tableShade}Weibull${}_2$ & \cellcolor{tableShade}0.065 & \cellcolor{tableShade}0.952 & \cellcolor{tableShade}0.034 & \cellcolor{tableShade}0.942 & \cellcolor{tableShade}0.018 & \cellcolor{tableShade}0.966 & \cellcolor{tableShade}0.598 & \cellcolor{tableShade}0.948 & \cellcolor{tableShade}0.201 & \cellcolor{tableShade}0.949 & \cellcolor{tableShade}0.201 & \cellcolor{tableShade}0.944\\ 
 & \cellcolor{white}KM & \cellcolor{white}0.065 & \cellcolor{white}0.957 & \cellcolor{white}0.034 & \cellcolor{white}0.945 & \cellcolor{white}0.018 & \cellcolor{white}0.941 & \cellcolor{white}0.599 & \cellcolor{white}0.931 & \cellcolor{white}0.200 & \cellcolor{white}0.932 & \cellcolor{white}0.202 & \cellcolor{white}0.948\\ 
\bottomrule
\end{tabular}
\caption{Simulation based on n=10,000 MZ and
  DZ twin pairs. Average (Av.) of estimates across 1,000 replications
  and coverage probabilities (Cv.) of corresponding 95\% confidence
  limits is shown for prevalence ($F_1$), MZ concordance
  ($\mathcal{C}_{\text{MZ}}$), DZ concordance
  ($\mathcal{C}_{\text{DZ}}$), and the variance components
  $\sigma_A^2$, $\sigma_C^2$ and $\sigma_E^2$. Results are shown for the
  naive estimator not taking the censoring into account (Naive), Weibull IPCW
  ignoring uncertainty in weights (Weibull${}_1$), Weibull IPCW with
  correct standard errors (Weibull${}_2$), and Kaplan-Meier without
  adjustment for uncertainty in weights (KM).
\label{tab:sim1}}

\end{sidewaystable}

The results of the simulation study are summarized in
Table~\ref{tab:sim1} with average estimates and coverage probabilities
of the 95\% confidence limits reported for the prevalence $F_1$,
concordance in MZ twins $\mathcal{C}_{\text{MZ}}$, concordance in DZ
twins $\mathcal{C}_{\text{DZ}}$, and the variance components
$\sigma_A^2$, $\sigma_C^2$, and $\sigma_E^2$.  In general the naive
estimates where the censoring mechanism is ignored shows very high
downward bias with poor coverage in both the prevalence and
concordance estimates, which is generally expected. In these
simulations the bias of the heritability estimate $\sigma_A^2$ is in
all cases negative with coverage that performs worse for larger true
value of $\sigma_A^2$. As discussed in \cite{scheike13:lida} the
direction of the bias in the heritability estimates may, however, go in
either direction depending on both the dependence structure and
censoring distribution.  The intuition for this is, that while the
concordance is biased downwards in both the MZ and DZ twins, it may change
relatively more/less in the DZ twins.

\newenvironment{DIFnomarkup}{}{}
\begin{DIFnomarkup}
\begin{table}
\footnotesize
\centering
\begin{tabular}{clrrrrrrrc}
\toprule
& & \multicolumn{1}{c}{True} & \multicolumn{3}{c}{IPCW} & \multicolumn{3}{c}{Naive} & \\ \cmidrule(lr){4-6}\cmidrule(lr){7-9}%%\cmidrule(lr){2-2}\cmidrule(lr){3-5}\cmidrule{6-8}
& & & Av. & Cov. & MSE & Av. & Cov. & MSE \\ \midrule
 &\cellcolor{col1}{$F_1$} &\cellcolor{col1}{ 0.065} &\cellcolor{col1}{ 0.065} &\cellcolor{col1}{ 0.962} &\cellcolor{col1}{0.0004} &\cellcolor{col1}{ 0.035} &\cellcolor{col1}{ 0.000} &\cellcolor{col1}{0.0916}\\ 
 &\cellcolor{col2}{$\rho_{MZ}$} &\cellcolor{col2}{ 0.667} &\cellcolor{col2}{ 0.664} &\cellcolor{col2}{ 0.970} &\cellcolor{col2}{0.0746} &\cellcolor{col2}{ 0.736} &\cellcolor{col2}{ 0.160} &\cellcolor{col2}{0.5282}\\ 
 &\cellcolor{col1}{$\rho_{DZ}$} &\cellcolor{col1}{ 0.500} &\cellcolor{col1}{ 0.499} &\cellcolor{col1}{ 0.951} &\cellcolor{col1}{0.1343} &\cellcolor{col1}{ 0.600} &\cellcolor{col1}{ 0.107} &\cellcolor{col1}{1.0914}\\ 
 &\cellcolor{col2}{$\mathcal{C}_{MZ}$} &\cellcolor{col2}{ 0.025} &\cellcolor{col2}{ 0.025} &\cellcolor{col2}{ 0.974} &\cellcolor{col2}{0.0003} &\cellcolor{col2}{ 0.014} &\cellcolor{col2}{ 0.000} &\cellcolor{col2}{0.0137}\\ 
 &\cellcolor{col1}{$\mathcal{C}_{DZ}$} &\cellcolor{col1}{ 0.018} &\cellcolor{col1}{ 0.018} &\cellcolor{col1}{ 0.951} &\cellcolor{col1}{0.0003} &\cellcolor{col1}{ 0.010} &\cellcolor{col1}{ 0.000} &\cellcolor{col1}{0.0065}\\ 
 &\cellcolor{col2}{$\lambda_{R,MZ}$} &\cellcolor{col2}{ 6.000} &\cellcolor{col2}{ 5.971} &\cellcolor{col2}{ 0.955} &\cellcolor{col2}{13.562} &\cellcolor{col2}{11.347} &\cellcolor{col2}{ 0.000} &\cellcolor{col2}{2897.3}\\ 
 &\cellcolor{col1}{$\lambda_{R,DZ}$} &\cellcolor{col1}{ 4.172} &\cellcolor{col1}{ 4.171} &\cellcolor{col1}{ 0.954} &\cellcolor{col1}{12.133} &\cellcolor{col1}{ 7.976} &\cellcolor{col1}{ 0.000} &\cellcolor{col1}{1484.2}\\ 
 &\cellcolor{col2}{$\log(\text{OR})_{MZ}$} &\cellcolor{col2}{ 2.670} &\cellcolor{col2}{ 2.660} &\cellcolor{col2}{ 0.968} &\cellcolor{col2}{1.7900} &\cellcolor{col2}{ 3.373} &\cellcolor{col2}{ 0.000} &\cellcolor{col2}{50.969}\\ 
 &\cellcolor{col1}{$\log(\text{OR})_{DZ}$} &\cellcolor{col1}{ 1.942} &\cellcolor{col1}{ 1.940} &\cellcolor{col1}{ 0.955} &\cellcolor{col1}{2.1320} &\cellcolor{col1}{ 2.662} &\cellcolor{col1}{ 0.004} &\cellcolor{col1}{53.658}\\ 
 &\cellcolor{col2}{$\sigma_A^2$} &\cellcolor{col2}{ 0.333} &\cellcolor{col2}{ 0.330} &\cellcolor{col2}{ 0.953} &\cellcolor{col2}{0.8823} &\cellcolor{col2}{ 0.272} &\cellcolor{col2}{ 0.866} &\cellcolor{col2}{0.9011}\\ 
 &\cellcolor{col1}{$\sigma_C^2$} &\cellcolor{col1}{ 0.333} &\cellcolor{col1}{ 0.334} &\cellcolor{col1}{ 0.946} &\cellcolor{col1}{0.6352} &\cellcolor{col1}{ 0.464} &\cellcolor{col1}{ 0.427} &\cellcolor{col1}{2.1052}\\ 
 &\cellcolor{col2}{$\sigma_E^2$} &\cellcolor{col2}{ 0.333} &\cellcolor{col2}{ 0.336} &\cellcolor{col2}{ 0.967} &\cellcolor{col2}{0.0746} &\cellcolor{col2}{ 0.264} &\cellcolor{col2}{ 0.137} &\cellcolor{col2}{0.5282}\\ \midrule 
 &\cellcolor{col1}{$F_1$} &\cellcolor{col1}{ 0.065} &\cellcolor{col1}{ 0.065} &\cellcolor{col1}{ 0.941} &\cellcolor{col1}{0.0005} &\cellcolor{col1}{ 0.035} &\cellcolor{col1}{ 0.000} &\cellcolor{col1}{0.0921}\\ 
 &\cellcolor{col2}{$\rho_{MZ}$} &\cellcolor{col2}{ 0.750} &\cellcolor{col2}{ 0.748} &\cellcolor{col2}{ 0.941} &\cellcolor{col2}{0.0618} &\cellcolor{col2}{ 0.804} &\cellcolor{col2}{ 0.213} &\cellcolor{col2}{0.3272}\\ 
 &\cellcolor{col1}{$\rho_{DZ}$} &\cellcolor{col1}{ 0.500} &\cellcolor{col1}{ 0.499} &\cellcolor{col1}{ 0.949} &\cellcolor{col1}{0.1396} &\cellcolor{col1}{ 0.601} &\cellcolor{col1}{ 0.108} &\cellcolor{col1}{1.0966}\\ 
 &\cellcolor{col2}{$\mathcal{C}_{MZ}$} &\cellcolor{col2}{ 0.030} &\cellcolor{col2}{ 0.030} &\cellcolor{col2}{ 0.948} &\cellcolor{col2}{0.0004} &\cellcolor{col2}{ 0.016} &\cellcolor{col2}{ 0.000} &\cellcolor{col2}{0.0196}\\ 
 &\cellcolor{col1}{$\mathcal{C}_{DZ}$} &\cellcolor{col1}{ 0.018} &\cellcolor{col1}{ 0.018} &\cellcolor{col1}{ 0.943} &\cellcolor{col1}{0.0003} &\cellcolor{col1}{ 0.010} &\cellcolor{col1}{ 0.000} &\cellcolor{col1}{0.0065}\\ 
 &\cellcolor{col2}{$\lambda_{R,MZ}$} &\cellcolor{col2}{ 7.166} &\cellcolor{col2}{ 7.154} &\cellcolor{col2}{ 0.944} &\cellcolor{col2}{17.438} &\cellcolor{col2}{13.565} &\cellcolor{col2}{ 0.000} &\cellcolor{col2}{4144.1}\\ 
 &\cellcolor{col1}{$\lambda_{R,DZ}$} &\cellcolor{col1}{ 4.172} &\cellcolor{col1}{ 4.173} &\cellcolor{col1}{ 0.947} &\cellcolor{col1}{13.063} &\cellcolor{col1}{ 7.996} &\cellcolor{col1}{ 0.000} &\cellcolor{col1}{1500.3}\\ 
 &\cellcolor{col2}{$\log(\text{OR})_{MZ}$} &\cellcolor{col2}{ 3.118} &\cellcolor{col2}{ 3.113} &\cellcolor{col2}{ 0.943} &\cellcolor{col2}{2.1903} &\cellcolor{col2}{ 3.824} &\cellcolor{col2}{ 0.000} &\cellcolor{col2}{51.544}\\ 
 &\cellcolor{col1}{$\log(\text{OR})_{DZ}$} &\cellcolor{col1}{ 1.942} &\cellcolor{col1}{ 1.939} &\cellcolor{col1}{ 0.945} &\cellcolor{col1}{2.2416} &\cellcolor{col1}{ 2.664} &\cellcolor{col1}{ 0.000} &\cellcolor{col1}{54.034}\\ 
 &\cellcolor{col2}{$\sigma_A^2$} &\cellcolor{col2}{ 0.500} &\cellcolor{col2}{ 0.499} &\cellcolor{col2}{ 0.945} &\cellcolor{col2}{0.8144} &\cellcolor{col2}{ 0.407} &\cellcolor{col2}{ 0.716} &\cellcolor{col2}{1.3176}\\ 
 &\cellcolor{col1}{$\sigma_C^2$} &\cellcolor{col1}{ 0.250} &\cellcolor{col1}{ 0.249} &\cellcolor{col1}{ 0.944} &\cellcolor{col1}{0.6247} &\cellcolor{col1}{ 0.397} &\cellcolor{col1}{ 0.332} &\cellcolor{col1}{2.5247}\\ 
 &\cellcolor{col2}{$\sigma_E^2$} &\cellcolor{col2}{ 0.250} &\cellcolor{col2}{ 0.252} &\cellcolor{col2}{ 0.938} &\cellcolor{col2}{0.0618} &\cellcolor{col2}{ 0.196} &\cellcolor{col2}{ 0.169} &\cellcolor{col2}{0.3272}\\ \midrule 
 &\cellcolor{col1}{$F_1$} &\cellcolor{col1}{ 0.065} &\cellcolor{col1}{ 0.065} &\cellcolor{col1}{ 0.952} &\cellcolor{col1}{0.0005} &\cellcolor{col1}{ 0.035} &\cellcolor{col1}{ 0.000} &\cellcolor{col1}{0.0919}\\ 
 &\cellcolor{col2}{$\rho_{MZ}$} &\cellcolor{col2}{ 0.800} &\cellcolor{col2}{ 0.799} &\cellcolor{col2}{ 0.949} &\cellcolor{col2}{0.0476} &\cellcolor{col2}{ 0.845} &\cellcolor{col2}{ 0.239} &\cellcolor{col2}{0.2205}\\ 
 &\cellcolor{col1}{$\rho_{DZ}$} &\cellcolor{col1}{ 0.500} &\cellcolor{col1}{ 0.499} &\cellcolor{col1}{ 0.955} &\cellcolor{col1}{0.1368} &\cellcolor{col1}{ 0.600} &\cellcolor{col1}{ 0.114} &\cellcolor{col1}{1.0871}\\ 
 &\cellcolor{col2}{$\mathcal{C}_{MZ}$} &\cellcolor{col2}{ 0.034} &\cellcolor{col2}{ 0.034} &\cellcolor{col2}{ 0.951} &\cellcolor{col2}{0.0005} &\cellcolor{col2}{ 0.018} &\cellcolor{col2}{ 0.000} &\cellcolor{col2}{0.0243}\\ 
 &\cellcolor{col1}{$\mathcal{C}_{DZ}$} &\cellcolor{col1}{ 0.018} &\cellcolor{col1}{ 0.018} &\cellcolor{col1}{ 0.939} &\cellcolor{col1}{0.0003} &\cellcolor{col1}{ 0.010} &\cellcolor{col1}{ 0.000} &\cellcolor{col1}{0.0065}\\ 
 &\cellcolor{col2}{$\lambda_{R,MZ}$} &\cellcolor{col2}{ 7.987} &\cellcolor{col2}{ 7.988} &\cellcolor{col2}{ 0.956} &\cellcolor{col2}{17.964} &\cellcolor{col2}{15.101} &\cellcolor{col2}{ 0.000} &\cellcolor{col2}{5109.9}\\ 
 &\cellcolor{col1}{$\lambda_{R,DZ}$} &\cellcolor{col1}{ 4.172} &\cellcolor{col1}{ 4.175} &\cellcolor{col1}{ 0.956} &\cellcolor{col1}{12.758} &\cellcolor{col1}{ 7.983} &\cellcolor{col1}{ 0.000} &\cellcolor{col1}{1489.1}\\ 
 &\cellcolor{col2}{$\log(\text{OR})_{MZ}$} &\cellcolor{col2}{ 3.441} &\cellcolor{col2}{ 3.442} &\cellcolor{col2}{ 0.951} &\cellcolor{col2}{2.3085} &\cellcolor{col2}{ 4.147} &\cellcolor{col2}{ 0.000} &\cellcolor{col2}{51.565}\\ 
 &\cellcolor{col1}{$\log(\text{OR})_{DZ}$} &\cellcolor{col1}{ 1.942} &\cellcolor{col1}{ 1.940} &\cellcolor{col1}{ 0.955} &\cellcolor{col1}{2.1908} &\cellcolor{col1}{ 2.662} &\cellcolor{col1}{ 0.000} &\cellcolor{col1}{53.664}\\ 
 &\cellcolor{col2}{$\sigma_A^2$} &\cellcolor{col2}{ 0.600} &\cellcolor{col2}{ 0.600} &\cellcolor{col2}{ 0.954} &\cellcolor{col2}{0.7214} &\cellcolor{col2}{ 0.489} &\cellcolor{col2}{ 0.596} &\cellcolor{col2}{1.6372}\\ 
 &\cellcolor{col1}{$\sigma_C^2$} &\cellcolor{col1}{ 0.200} &\cellcolor{col1}{ 0.199} &\cellcolor{col1}{ 0.958} &\cellcolor{col1}{0.5866} &\cellcolor{col1}{ 0.356} &\cellcolor{col1}{ 0.262} &\cellcolor{col1}{2.7722}\\ 
 &\cellcolor{col2}{$\sigma_E^2$} &\cellcolor{col2}{ 0.200} &\cellcolor{col2}{ 0.201} &\cellcolor{col2}{ 0.945} &\cellcolor{col2}{0.0476} &\cellcolor{col2}{ 0.155} &\cellcolor{col2}{ 0.178} &\cellcolor{col2}{0.2205}
\\ 
\bottomrule
\end{tabular}
\caption{Simulation based on n=10,000 MZ and
  DZ twin pairs with continuous covariate affecting both the
  censoring mechanism and the transition probabilities to cancer and death\label{tab:simx}.
  Average (Av.) of estimates across 1,000 replications,
  coverage probabilities (Cv.) of corresponding 95\% confidence
  limits, and Mean Squared Error multiplied by 100 (MSE)
  is shown for prevalence ($F_1$), concordance
  ($\mathcal{C}_{\text{MZ}}$, $\mathcal{C}_{\text{DZ}}$), 
  relative recurrence risks ratios ($\lambda_{R,\text{MZ}}$,
  $\lambda_{R,\text{DZ}}$), and log odds-ratios
  ($\log(\text{OR})_{MZ}$, $\log(\text{OR})_{DZ}$),
  and the variance components
  $\sigma_A^2$, $\sigma_C^2$ and $\sigma_E^2$. 
  Results are shown for the
  naive estimator ignoring the censorings (Naive), and Weibull IPCW
  using a correct model for the censoring (IPCW).
}
\end{table}
\end{DIFnomarkup}

Generally, the loss in efficiency using the Kaplan-Meier estimator
seemed to be very modest. Interestingly, the two IPCW-adjusted
estimators ignoring the uncertainty in the estimated weights (KM and
Weibull${}_1$) showed excellent coverage probabilities across almost
all scenarios. This may be explained by the high degree of censoring
(as also seen in the real data), which causes the variance of the
estimator to be dominated
%%by may cause near orthogonality between the data-set
%%used to estimate the censoring distribution and in 
by the variance of the estimator based
on \eqref{eq:ipcw1} where only the uncensored pairs are used.  A
tendency was in fact seen towards slightly smaller coverage when the
censoring was smaller and heritability higher, while the estimator
with confidence limits based on \eqref{eq:IF3} performs seemingly
better here.  Ignoring the estimated censoring probabilities can in
some situations lead to conservative estimates
\citep{rotnitzkyrobbins1995}. This does not seem to be the case here,
and may be a consequence of the estimator of the censoring distribution
being a GEE-type estimator and not a MLE.

We also examined the effect of introducing a covariate affecting both
the censoring and transition probabilities to death or cancer. Given a
normal distributed covariate $X\sim\mathcal{N}(0,0.25)$, the shared
environmental effect (C component) was defined as 
$\eta^C=0.5X+\eta^{C_0}$ with $\eta^{C_0}\sim\mathcal{N}(0,\sigma^2_C
- 0.0625)$, and the random effect for the competing risk of death was
defined as $\eta_2=\eta_C-0.25X$.  For the censoring mechanism we used
a proportional hazards model with baseline hazard as described
above, with shape-parameter $\log(\nu)=0.5$, such that the cumulative
hazard took the form $\Lambda(t) = \Lambda_0(t)\exp(-X)$. Results are
summarized in Table \ref{tab:simx}, and are generally very comparable
to the results of Table \ref{tab:sim1}.

% distribution(m,~c) <- weibull.lvm(shape=1/exp(logshape),scale=-logscale)
% distribution(m,~x) <- X
% dc <- sim(m,length(X),p=c("c~x"=beta))

% -----
% We also examined the convergence of the $\wAIC$... See Table~\ref{tab:simaic}

%  AIC is quite
% variable, however within the expected Monte Carlo variation.
% Table~\ref{tab:simaic}.
% \begin{table}
%   \begin{tabular}{ccc}
%     \toprule
%     \bottomrule
%   \end{tabular}
%   \caption{Simulation based on 10,000 replications with n=10,000 MZ and
%     DZ twin pairs.\label{tab:simaic}}
% \end{table}

%}}} Simulation
 
%{{{ Application prostate

\section{Application to twin cancer data}\label{sec:application}

Studying genetic influence on the complex trait of cancer is central
in themes of etiology, treatment and prevention. 
% Extensive search for genetic variants has led to some insight and
% variants in linkage has been reported (\cite{GWASprostate}).
Twin and general family studies have reported low to moderate genetic
influence (\cite{lichtenstein2000environmental} and
\cite{baker2005biometrics}).  Based on a combined Nordic study of the
Danish, Finnish and Swedish twins registries,
\cite{lichtenstein2000environmental} concluded that 42\% of variation
in prostate-cancer liability was due to genetic factors (95\%
confidence limits 0.29--0.50).  However, in these cohorts around 70\%
of the participants are censored resulting in biased estimates of all
population parameters including prevalences, concordances
rates and heritability, as discussed in the previous sections.

%\section{Biometric modelling of twin data}
%%% FIX 58 \% was due to non-shared environmental factors and residual
%%% variation

We investigate genetic influence on prostate cancer using the
population based twin cohort of Danish twins born 1900 to 1982
constituting N = $15,509$ male pairs of whom $5,488$ MZ and $10,021$
same sex male DZ pairs are eligible for studying prostate cancer. The
cohort is followed up with respect to survival status as of July 
2009. Data on cancer diagnosis, status and time of event, were
obtained from the National Cancer Registry which was initiated in 1943
(See \cite{hjelmborgprostate} for further
description of the cohort). Numbers of pairs by status of cancer and death
can be seen in Table~\ref{tab:pairs}.

\begin{table}[htbp]\centering
%\textbf{Number of pairs at time of follow-up}
  \begin{tabular}{@{}  l l l  l  @{}}
    \toprule
    \multicolumn{4}{@{}c@{}}{\textbf{Number of pairs at time of
        follow-up}} \\
    \hline
MZ \& DZ & Prostate cancer & No cancer and dead & No cancer and alive
\\ \midrule
  Prostate cancer & 25 \& 14 & 178 & 108 \\ 
  No cancer and dead & 70 & 843 \& 1,694 & 1,319 \\ 
  No cancer and alive & 39 & 492 & 4,019 \& 6,708 \\ 
  \bottomrule
  \end{tabular}
\caption{Number of pairs by status at time of follow-up with MZ pairs in lower left triangle and DZ pairs in upper right triangle. }
\label{tab:pairs}
\end{table}

There was a significant difference in the censoring distributions in
MZ and DZ twins. This may in part be explained by increased used of In
Vitro Fertilizations over time which have caused a chance in DZ/MZ
distribution and perhaps consequently also censoring distributions in
this cohort. We therefore based the IPCW-model on a stratified
Kaplan-Meier model.

As described in Section~\ref{sec:modelselect} we first examined if the
marginal distributions within MZ and DZ twins could be assumed to be
the same (p=0.52). In the reduced model with identical marginals the
tetrachoric correlations was 0.63 (0.47--0.75) for MZ pairs and 0.25
(0.07--0.41) for DZ pairs. A test for genetic effects was performed by
comparing these correlation coefficients which yielded a p-value of
p=0.001, indicating strong evidence in support of a genetic
contribution. In the polygenic models the \wAIC\, was slightly in favour
of the ADE model but very similar results was obtained from AE and ACE models
in terms of broad-sense heritability.  For the chosen ADE model the
broad-sense heritability was 0.63 (0.49--0.77). The results are summarized
in Table~\ref{tab:prostateresults} together with the biased naive
estimates as a reference. Here we also report the \textit{casewise
  concordance} \citep{witte99:_likel_based_approac_estim_twin}, i.e.,
the conditional probability that a twin gets cancer given the co-twin
got cancer, and the \textit{relative recurrence risk ratio} which
describes the excess risk of prostate cancer for a twin given the
co-twin got prostate cancer, compared to the marginal (population)
risk
\begin{align*}
    \lambda_R = \frac{\pr(T_1\leq\tau, T_2\leq\tau, \cause_1=1,\cause_2=1)}{\pr(T_1\leq\tau,\cause_1=1)^2}.
\end{align*}
All estimates except for the heritability are reported from the more
parsimonious bivariate Probit model but results was almost identical
with the estimates from the ADE-model. 

In conclusion, we see strong evidence for a genetic component in the
development of prostate cancer. As expected,
the naive estimator provides heavily downward biased estimates of the
prevalence and concordance, and in this case upward bias of the
heritability estimates and relative recurrence risk ratio estimates.

\begin{table}[htbp]
  \centering
  \begin{tabular}{lrr}
    \toprule
    & \textbf{IPCW-adjusted} & \textbf{Naive} \\
    \toprule
$F_1$ & 0.055 (0.049; 0.062) & 0.015 (0.014; 0.017) \\
%%$F_1^2$ & 0.003 (0.002; 0.004) &  0.00023 (0.00019; 0.00028) \\ %%$2.3\cdot 10^{-4} (1.9; 2.8)\cdot 10^{-4}$ \\ 
\midrule
$\rho_{\text{MZ}}$ & 0.626 (0.466; 0.746) & 0.730 (0.629; 0.807) \\
$\rho_{\text{DZ}}$ & 0.248 (0.068; 0.412) & 0.350 (0.224; 0.465) \\
\midrule
$\mathcal{C}_{\text{MZ}}$ & 0.019 (0.013; 0.027) & 0.005 (0.004; 0.007) \\
$\mathcal{C}_{\text{DZ}}$ & 0.007 (0.004; 0.013) & 0.001 (0.001;
0.002) \\
\midrule
$\mathcal{C}_{\text{MZ}}/F_1$ & 0.340 (0.241; 0.455) & 0.324 (0.240; 0.421) \\
$\mathcal{C}_{\text{DZ}}/F_1$ & 0.130 (0.076; 0.215) & 0.087 (0.053; 0.140) \\
\midrule
$\lambda_{R,{\text{MZ}}}$ & 6.166 (4.132; 8.201) & 21.17 (15.25; 27.10) \\
$\lambda_{R,{\text{DZ}}}$ & 2.360 (1.148; 3.571) & 5.713 (2.966; 8.459)
\\
\midrule
$H^2$ & 0.626 (0.486; 0.766) & 0.73 (0.642; 0.819) \\
    \bottomrule    
  \end{tabular}
  \caption{Estimates (and 95\% confidence limits) of association of prostate cancer for MZ and DZ
    twins based on bivariate Probit model. The first column
    contains the IPCW-adjusted estimates and the second column the
    biased estimates ignoring the right-censoring mechanism. 
    We show estimates of prevalence $F_1$,
    tetrachoric correlations $\rho$, concordance $\mathcal{C}$,
    casewise concordance $\mathcal{C}/F_1$, and relative recurrence
    risk ratio $\lambda_R$. The broad-sense heritability estimate $H^2$ is based on an ADE-model.}
  \label{tab:prostateresults}
\end{table}

\begin{figure}
  \centering
  \mbox{
    \includegraphics[width=0.49\textwidth]{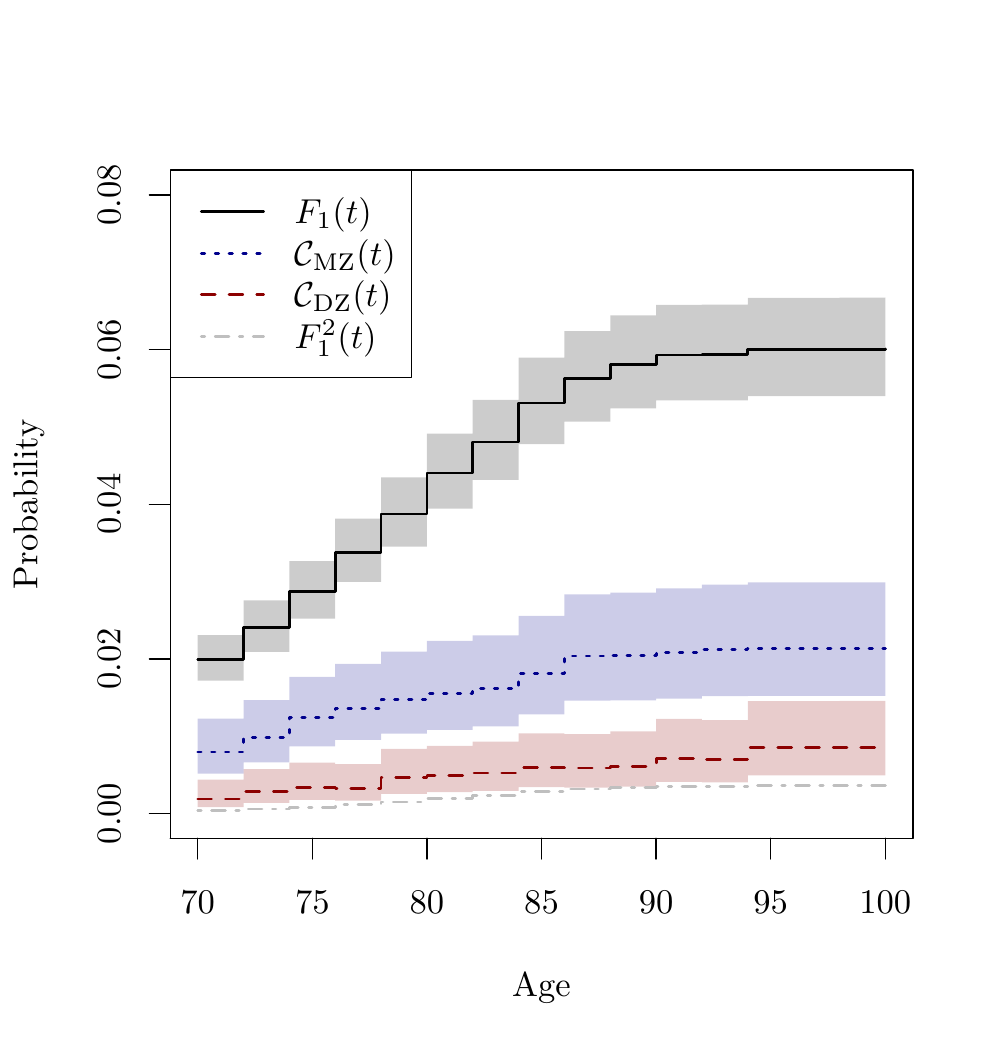}
    \includegraphics[width=0.49\textwidth]{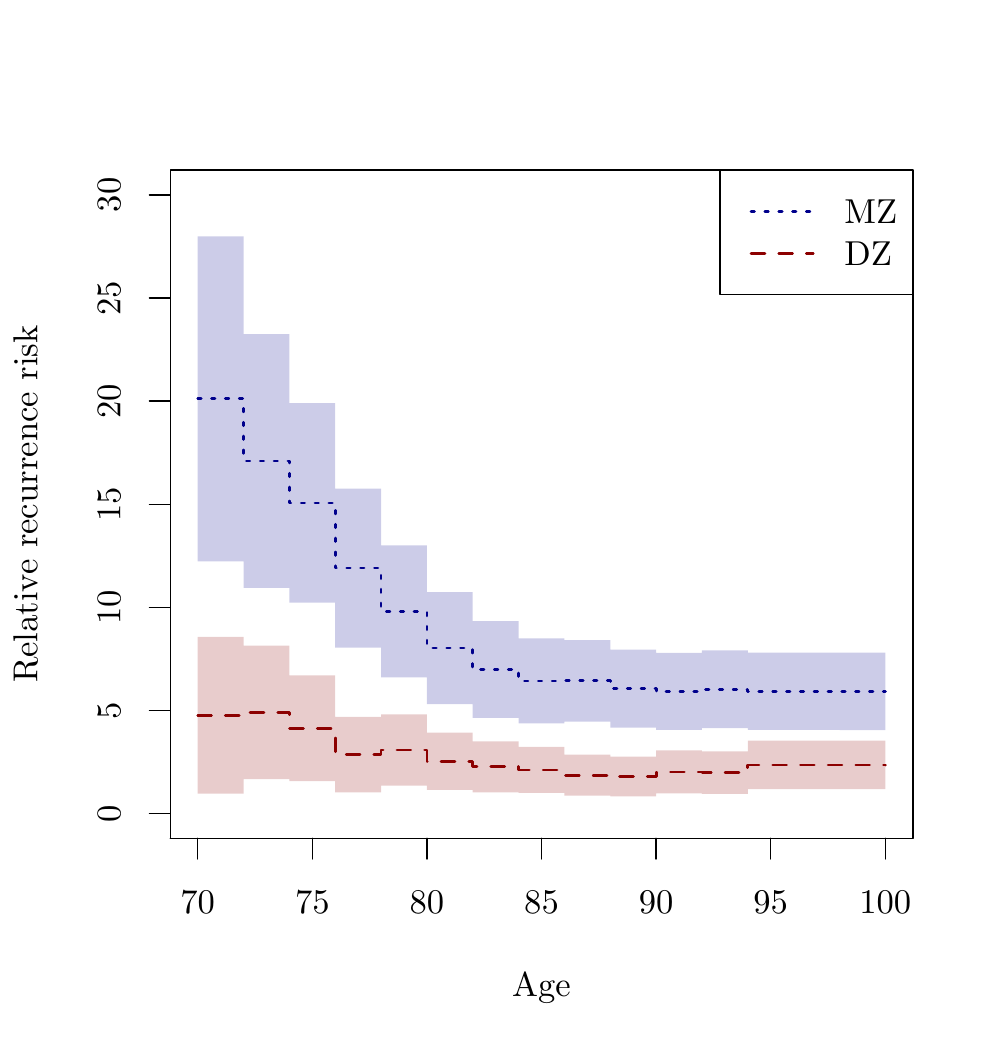}}
  \caption{Concordance and relative recurrence risk ratio estimates
    for prostate cancer in MZ and DZ twins. The left panel shows the
    concordance for prostate cancer in MZ and DZ twins with point-wise
    95\% confidence limits calculated at different ages in two-years
    intervals. The two concordance functions are bounded above by the
    marginal cumulative incidence corresponding to perfect dependence
    and below by the squared marginal corresponding to
    independence. In the right panel the relative recurrence risk
    ratio is shown for MZ and DZ twins for different ages with
    point-wise 95\% confidence limits.\label{fig:timerisk}}
\end{figure}

\begin{figure}
  \centering
    \includegraphics[width=0.7\textwidth]{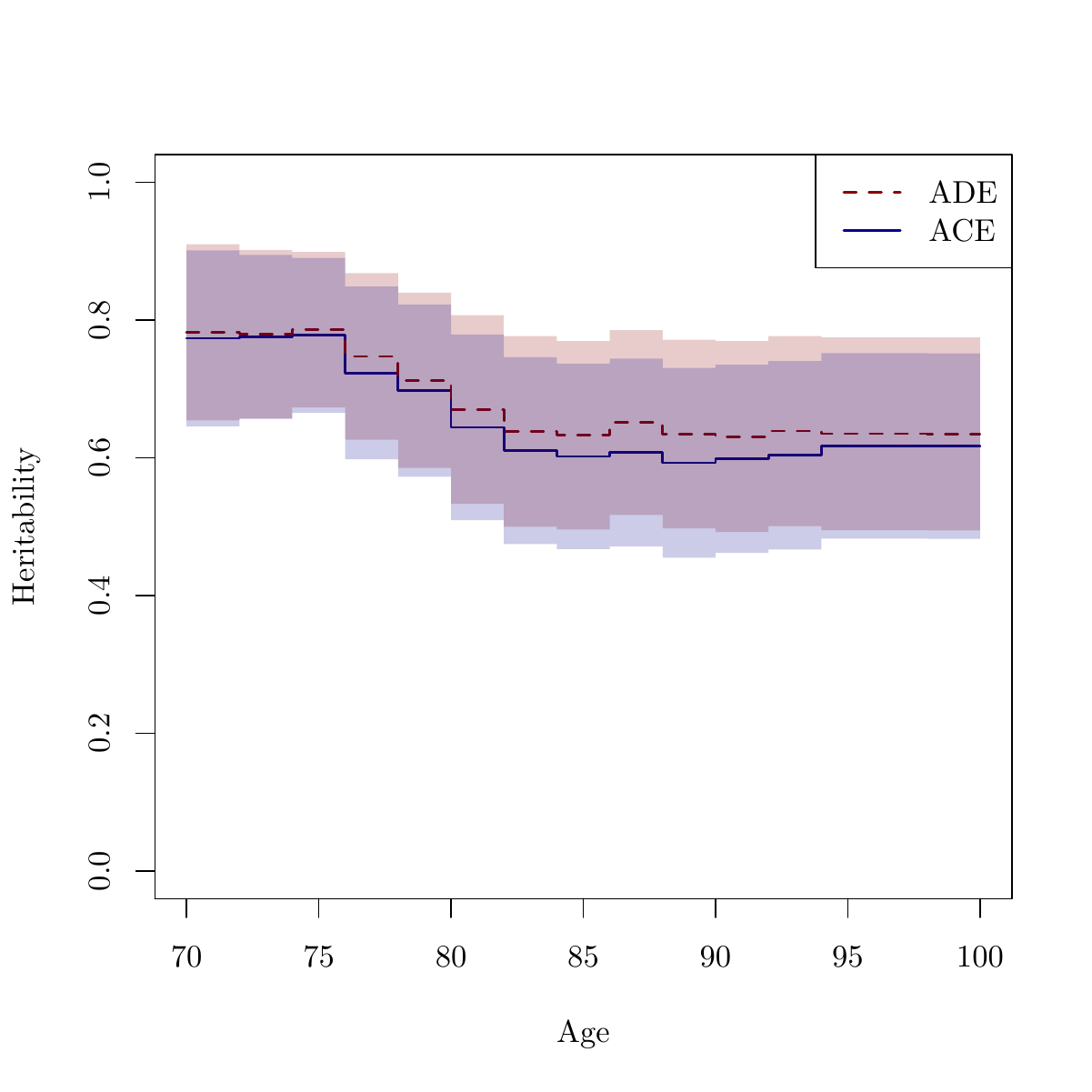}
    \caption{Heritability of prostate cancer calculated at different
      ages in two years intervals with point-wise 95\% confidence
      limits. Estimates are based on IPCW adjusted ACE (solid line)
      and ADE (dashed line) models.\label{fig:timeher}}
\end{figure}

We also examined how the association between MZ and DZ twins might
depend on age by choosing different values of $\tau$ in
\eqref{eq:tau}, with different parameters at each time point.  As
shown in Figure \ref{fig:timerisk}, this allows us to describe the
cumulative incidence function for prostate cancer and the concordance
functions and relative recurrence risk ratios as functions of age
based on the flexible bivariate Probit model. We also calculated the
heritability for both an ACE and ADE model (Figure~\ref{fig:timeher})
which in agreement indicates higher genetic contribution in earlier
ages. This stronger dependence in early-onset has been suggested for
several types of cancers.

%}}} Application

%{{{ Discussion
\section{Discussion}\label{sec:discussion}

There has been considerable interest in quantifying the genetic
influence of cancer, and family and twin studies have here served as
important tools. The censoring problem we have discussed in this paper
seems to have been largely ignored in the epidemiological literature,
which makes estimates from these studies difficult to interpret.

We have here presented a simple method based on inverse probability
weighting that corrects for a major source of bias by taking advantage
of the time to event information that is most often provided in cohort
studies along with the binary disease status. The method allows for
flexible and computational robust modelling of twin dependence at
different ages. Our simulations show that the method performs very
well in a realistic setup.

Applied on data from the Danish Twin Registry and Danish Cancer
Registry we estimated a heritability of 0.62 in prostate cancer, and
relative recurrence risk ratios of 6.2 in MZ twins and 2.4 in DZ twins.

\vspace*{\bigskipamount}

Here we have only considered twins but both the estimation and
computational framework can be generalized to larger pedigrees. Also,
extensions to ascertained samples should follow along the lines of
\cite{javarasACEcasecontrol2010}. Another topic for future research
will be development of efficient and robust estimating equations.

% On a note a
% slightly more efficient estimator could be obtained by also using the
% singletons (observations where only one twin is censored) in the
% analysis. In practice this can be done by double-entering each twin
% pair, such that every pair is entering the dataset twice with the role
% of twin 1 and twin 2 switched, and using the estimating equation
% \begin{align*}
% \mathcal{U}(\theta) = \frac{2\Delta_{i1}\Delta_{i2}}{G_c(t_1,t_2)}\mathcal{U}(\theta)
% + \frac{\Delta_{i1}}{G_c(t_1)}\mathcal{U}_{marg}(\theta) = 0.
% \end{align*}
% % GEE-type estimating equation where terms are aggregated to form an
% % iid decomposition (influence curves).
% Consistency of this double-entering follows from consistency of both
% marginal and bivariate censoring.

% Some controversy exists regarding heritability estimates [references...]
% In fact just a simple measure of difference in association so in that
% sense we believe it is a useful measure...

% \item Note left-truncation in application. To be pursued.
% \item robustness (augmented versions)
% \item Same censoring.
% \item Covariates
% \end{itemize}

All methods are available in the R package \texttt{mets}
\citep{metspackage}.

\section*{Acknowledgement}
We thank our collaborators at the NorTwinCan consortia, that received
support from a Nordic Cancer Union grant and the Ellison Foundation.
%}}} Discussion

%{{{ Endmatter

%\bibliographystyle{elsarticle-num}
\bibliographystyle{elsarticle-harv}
\bibliography{bptwin} 

\begin{thebibliography}{34}
\expandafter\ifx\csname natexlab\endcsname\relax\def\natexlab#1{#1}\fi
\expandafter\ifx\csname url\endcsname\relax
  \def\url#1{\texttt{#1}}\fi
\expandafter\ifx\csname urlprefix\endcsname\relax\def\urlprefix{URL }\fi

\bibitem[{Aalen and Johansen(1978)}]{aale:joha:1978}
Aalen, O.~O., Johansen, S., 1978. An empirical transition matrix for
  non-homogeneous {M}arkov chains based on censored observations. Scandinavian
  Journal of Statistics 5, 141--150.

\bibitem[{Akaike(1973)}]{Akaike73}
Akaike, H., 1973. Information theory and an extension of the maximum likelihood
  principle. In: Petrov, B.~N., Csaki, F. (Eds.), Second International
  Symposium on Information Theory. Budapest: Akademiai Kiado, pp. 267--281.

\bibitem[{Andersen et~al.(1993)Andersen, Borgan, Gill, and
  Keiding}]{andersencounting}
Andersen, P.~K., Borgan, O., Gill, R.~D., Keiding, N., 1993. {Statistical
  Models Based on Counting Processes (Springer Series in Statistics)}.
  Springer.

\bibitem[{Ashford and Sowden(1970)}]{ashford70probit}
Ashford, J., Sowden, R., 1970. Multivariate probit analysis. Biometrics 26,
  535--546.

\bibitem[{Baker et~al.(2005)Baker, Lichtenstein, Kaprio, and
  Holm}]{baker2005biometrics}
Baker, S.~G., Lichtenstein, P., Kaprio, J., Holm, N., 2005. Genetic
  susceptibility to prostate, breast, and colorectal cancer among nordic twins.
  Biometrics 61~(1), 55--63.
\newline\urlprefix\url{http://dx.doi.org/10.1111/j.0006-341X.2005.030924.x}

\bibitem[{Duncan(2004)}]{duncan04:genepi}
Duncan, T.~C., 2004. Statistical methods in genetic epidemiology. Oxford
  Universitet Press, New York.

\bibitem[{Falconer(1967)}]{falconer67}
Falconer, D.~S., Aug 1967. {{T}he inheritance of liability to diseases with
  variable age of onset, with particular reference to diabetes mellitus}. Ann.
  Hum. Genet. 31~(1), 1--20.
\newline\urlprefix\url{http://dx.doi.org/10.1111/j.1469-1809.1967.tb01249.x}

\bibitem[{Falconer and Mackay(1996)}]{falconer-mackay-1994}
Falconer, D.~S., Mackay, T. F.~C., 1996. {Introduction to Quantitative
  Genetics}, 4th Edition. {Prentice Hall}.

\bibitem[{Fine and Gray(1999)}]{fine_and_gray_competing_risk_1999}
Fine, J.~P., Gray, R.~J., 1999. A proportional hazards model for the
  subdistribution of a competing risk. Journal of the American Statistical
  Association 94~(446), 496--509.

\bibitem[{Fisher(1918)}]{Fisher:1918}
Fisher, R.~A., 1918. The correlation between relatives on the supposition of
  mendelian inheritance. Transactions of the Royal Society of Edinburgh 52,
  399--433.
\newline\urlprefix\url{http://dx.doi.org/10.1017/S0080456800012163}

\bibitem[{Genz(1992)}]{genz92:_numer_comput_of_multiv_normal_probab}
Genz, A., 1992. Numerical computation of multivariate normal probabilities.
  Journal of Computational and Graphical Statistics 1~(2), 141--149.
\newline\urlprefix\url{http://dx.doi.org/10.1080/10618600.1992.10477010}

\bibitem[{Gill(1980)}]{gill-thesis}
Gill, R.~D., 1980. Censoring and stochastic integrals. Ph.D. thesis,
  Matematisch Centrum, Amsterdam.

\bibitem[{Gorfine and Hsu(2011)}]{gorfinehsu2011}
Gorfine, M., Hsu, L., 2011. Frailty-based competing risks model for
  multivariate survival data. Biometrics 67~(2), 415--426.
\newline\urlprefix\url{http://dx.doi.org/10.1111/j.1541-0420.2010.01470.x}

\bibitem[{Hjelmborg et~al.(2014)Hjelmborg, Scheike, K.Holst, Skytthe, Penney,
  Graff, Pukkala, Christensen, Adami, Holm, Hansen, Hartman, Czene, Harris,
  Kaprio, and Mucci}]{hjelmborgprostate}
Hjelmborg, J., Scheike, T., K.Holst, K., Skytthe, A., Penney, K.~L., Graff,
  R.~E., Pukkala, E., Christensen, K., Adami, H.-O., Holm, N., Hansen, S.,
  Hartman, M., Czene, K., Harris, J.~R., Kaprio, J., Mucci, L.~A., 2014. The
  heritability of prostate cancer in the nordic twin study of cancer. Cancer
  Epidemiology, Biomarkers \& PreventionIn press.

\bibitem[{Holst et~al.(2011)Holst, Budtz-J{\o}rgensen, and
  Knudsen}]{holst:binarylatent}
Holst, K.~K., Budtz-J{\o}rgensen, E., Knudsen, G.~M., 2011. A latent variable
  model with mixed binary and continuous response variables. Tech. Rep.~5,
  University of Copenhagen, Department of Biostatistics.

\bibitem[{Holst and Scheike(2014)}]{metspackage}
Holst, K.~K., Scheike, T., 2014. mets (R package). Version 0.2.8.
\newline\urlprefix\url{http://cran.r-project.org/web/packages/mets/}

\bibitem[{Horvitz and Thompson(1952)}]{horvitzthompson1952}
Horvitz, D.~G., Thompson, D.~J., 1952. A generalization of sampling without
  replacement from a finite universe. Journal of the American Statistical
  Association 47~(260), 663--685.
\newline\urlprefix\url{http://dx.doi.org/10.1080/01621459.1952.10483446}

\bibitem[{Javaras et~al.(2010)Javaras, Hudson, and
  Laird}]{javarasACEcasecontrol2010}
Javaras, K.~N., Hudson, J.~I., Laird, N.~M., 2010. Fitting ace structural
  equation models to case-control family data. Genetic Epidemiology 34~(3),
  238--245.
\newline\urlprefix\url{http://dx.doi.org/10.1002/gepi.20454}

\bibitem[{Lange(1997)}]{lange97}
Lange, K., 1997. An approximate model of polygenic inheritance. Genetics
  147~(3), 1423--1430.
\newline\urlprefix\url{http://www.genetics.org/content/147/3/1423.abstract}

\bibitem[{Lange(2002)}]{lange02:_mathem_statis_method_genet_analy}
Lange, K., 2002. Mathematical and Statistical Methods for Genetic Analysis,
  second edition Edition. Springer-Verlag New York.

\bibitem[{Lichtenstein et~al.(2000)Lichtenstein, Holm, Iliadou, Kaprio,
  Koskenvuo, Pukkala, Skytthe, and Hemminki}]{lichtenstein2000environmental}
Lichtenstein, P., Holm, N.~V., Iliadou, P. K. V.~A., Kaprio, J., Koskenvuo, M.,
  Pukkala, E., Skytthe, A., Hemminki, K., 2000. {E}nvironmental and {H}eritable
  {F}actors in the {C}ausation of {C}ancer. {A}nalyses of {C}ohorts of {T}wins
  from {S}weden, {D}enmark, and {F}inland. The New England Journal of Medicine
  343~(2), 78--85.
\newline\urlprefix\url{http://dx.doi.org/10.1056/NEJM200007133430201}

\bibitem[{Lin(2000)}]{linmedicalcost2000}
Lin, D., 2000. Linear regression analysis of censored medical costs.
  Biostatistics 1~(1), 35--47.
\newline\urlprefix\url{http://dx.doi.org/10.1093/biostatistics/1.1.35}

\bibitem[{Neale and Cardon(1992)}]{nealecardon}
Neale, M., Cardon, L., 1992. Methodology for Genetic Studies of Twins and
  Families. Kluwer Academic Publishers, Dordrecht, Netherlands.

\bibitem[{Pan(2001)}]{QICpan01}
Pan, W., 2001. Akaike's information criterion in generalized estimating
  equations. Biometrics 57~(1), 120--125.
\newline\urlprefix\url{http://dx.doi.org/10.1111/j.0006-341X.2001.00120.x}

\bibitem[{Ripatti et~al.(2003)Ripatti, Gatz, Pedersen, and
  Palmgren}]{ripatti2003}
Ripatti, S., Gatz, M., Pedersen, N.~L., Palmgren, J., 2003. Three-state frailty
  model for age at onset of dementia and death in swedish twins. Genetic
  Epidemiology 24~(2), 139--149.
\newline\urlprefix\url{http://dx.doi.org/10.1002/gepi.10209}

\bibitem[{Robins and Rotnitzky(1992)}]{robins92}
Robins, J.~M., Rotnitzky, A., 1992. Recovery of information and adjustment for
  dependent censoring using surrogate markers. In: Jewell, N.~P., Dietz, K.,
  Farewell, V.~T. (Eds.), AIDS Epidemiology. Birkhauser Boston, pp. 297--331.
\newline\urlprefix\url{http://dx.doi.org/10.1007/978-1-4757-1229-2\_14}

\bibitem[{Rotnitzky and Robins(1995)}]{rotnitzkyrobbins1995}
Rotnitzky, A., Robins, J.~M., 1995. Semiparametric regression estimation in the
  presence of dependent censoring. Biometrika 82~(4), 805--820.

\bibitem[{Scheike et~al.(2014{\natexlab{a}})Scheike, Holst, and
  Hjelmborg}]{scheike13:lida}
Scheike, T.~H., Holst, K.~K., Hjelmborg, J.~B., 2014{\natexlab{a}}. Estimating
  heritability for cause specific mortality based on twin studies. Lifetime
  Data Analysis 20~(2), 210--233.
\newline\urlprefix\url{http://dx.doi.org/10.1007/s10985-013-9244-x}

\bibitem[{Scheike et~al.(2014{\natexlab{b}})Scheike, Holst, and
  Hjelmborg}]{scheike13:concordance}
Scheike, T.~H., Holst, K.~K., Hjelmborg, J.~B., 2014{\natexlab{b}}. Estimating
  twin concordance for bivariate competing risks twin data. Statistics in
  Medicine 33~(7), 1193--1204.
\newline\urlprefix\url{http://dx.doi.org/10.1002/sim.6016}

\bibitem[{Scheike et~al.(2008)Scheike, Zhang, and
  Gerds}]{scheikebinomialregr08}
Scheike, T.~H., Zhang, M.-J., Gerds, T.~A., 2008. Predicting cumulative
  incidence probability by direct binomial regression. Biometrika 95~(1),
  205--220.
\newline\urlprefix\url{http://dx.doi.org/10.1093/biomet/asm096}

\bibitem[{Sham(1998)}]{shamhuman}
Sham, P., 1998. Statistics in Human Genetics. Arnold Appl. of Statistics.

\bibitem[{Stefanski and Boos(2002)}]{stefanski_boos_2002_m-estimator}
Stefanski, L.~A., Boos, D.~D., 2002. The calculus of m-estimation. The American
  Statistician 56~(1), 29--38.

\bibitem[{Tsiatis(2006)}]{tsiatis2006semiparametric}
Tsiatis, A., 2006. Semiparametric Theory and Missing Data. Springer Series in
  Statistics. Springer New York.

\bibitem[{Witte et~al.(1999)Witte, Carlin, and
  Hopper}]{witte99:_likel_based_approac_estim_twin}
Witte, J.~S., Carlin, J.~B., Hopper, J.~L., 1999. Likelihood-based approach to
  estimating twin concordance for dichotomous traits. Genetic Epidemiology 16,
  290--304.

\end{thebibliography}

\end{document}